\RequirePackage{latexrelease}
\documentclass[a4paper,twocolumn,11pt,unpublished]{quantumarticle}
\pdfoutput=1
\usepackage[english]{babel}
\usepackage[T1]{fontenc}
\usepackage{amsmath}
\usepackage{hyperref}

\usepackage{graphicx}
\usepackage{qcircuit}
\usepackage{multirow}

\begin{document}

\title{Simulating large-size quantum spin chains on cloud-based superconducting quantum computers}
\author{Hongye Yu}
\affiliation{Department of Physics and Astronomy, State University of New York at
Stony Brook, Stony Brook, NY 11794-3800, USA}
\author{Yusheng Zhao}
\affiliation{Department of Physics and Astronomy, State University of New York at
Stony Brook, Stony Brook, NY 11794-3800, USA}
\author{Tzu-Chieh Wei}
\affiliation{Department of Physics and Astronomy, State University of New York at
Stony Brook, Stony Brook, NY 11794-3800, USA}
\affiliation{C. N. Yang Institute for Theoretical Physics,
State University of New York at
Stony Brook, Stony Brook, NY 11794-3840, USA}
\affiliation{Institute for Advanced Computational Science,
State University of New York at
Stony Brook, Stony Brook, NY 11794-5250, USA}

\begin{abstract}
Quantum computers have the potential to efficiently simulate large-scale quantum systems for which classical approaches are bound to fail. 
 Even though several existing quantum devices now feature total qubit numbers of more than one hundred, their applicability remains plagued by the presence of noise and errors. Thus, the  degree to which large quantum systems can successfully be simulated on these devices remains unclear. Here, we report on cloud simulations performed on several of IBM's superconducting quantum computers to simulate ground states of spin chains having a wide range of system sizes up to one hundred and two qubits. We find that the ground-state energies extracted from realizations  across different quantum computers and  system sizes reach the expected values to within errors that are small (i.e. on the percent level), including the inference of the energy density in the thermodynamic limit from these values.     We achieve this accuracy through a combination of physics-motivated variational Ansatzes, and efficient, scalable energy-measurement and error-mitigation protocols, including the use of a reference state in the zero-noise extrapolation. By using a 102-qubit system, we have been able to successfully apply up to 3186 CNOT gates in a single circuit when performing gate-error mitigation. Our accurate, error-mitigated results for random parameters in the Ansatz states suggest that a standalone hybrid quantum-classical variational approach for large-scale XXZ models is feasible.
\end{abstract}

\section{Introduction.}
\begin{figure*}[ht!]
\begin{tabular}{ll}
  (a)  & (b)  \\ 
  \includegraphics[width=0.48\textwidth]{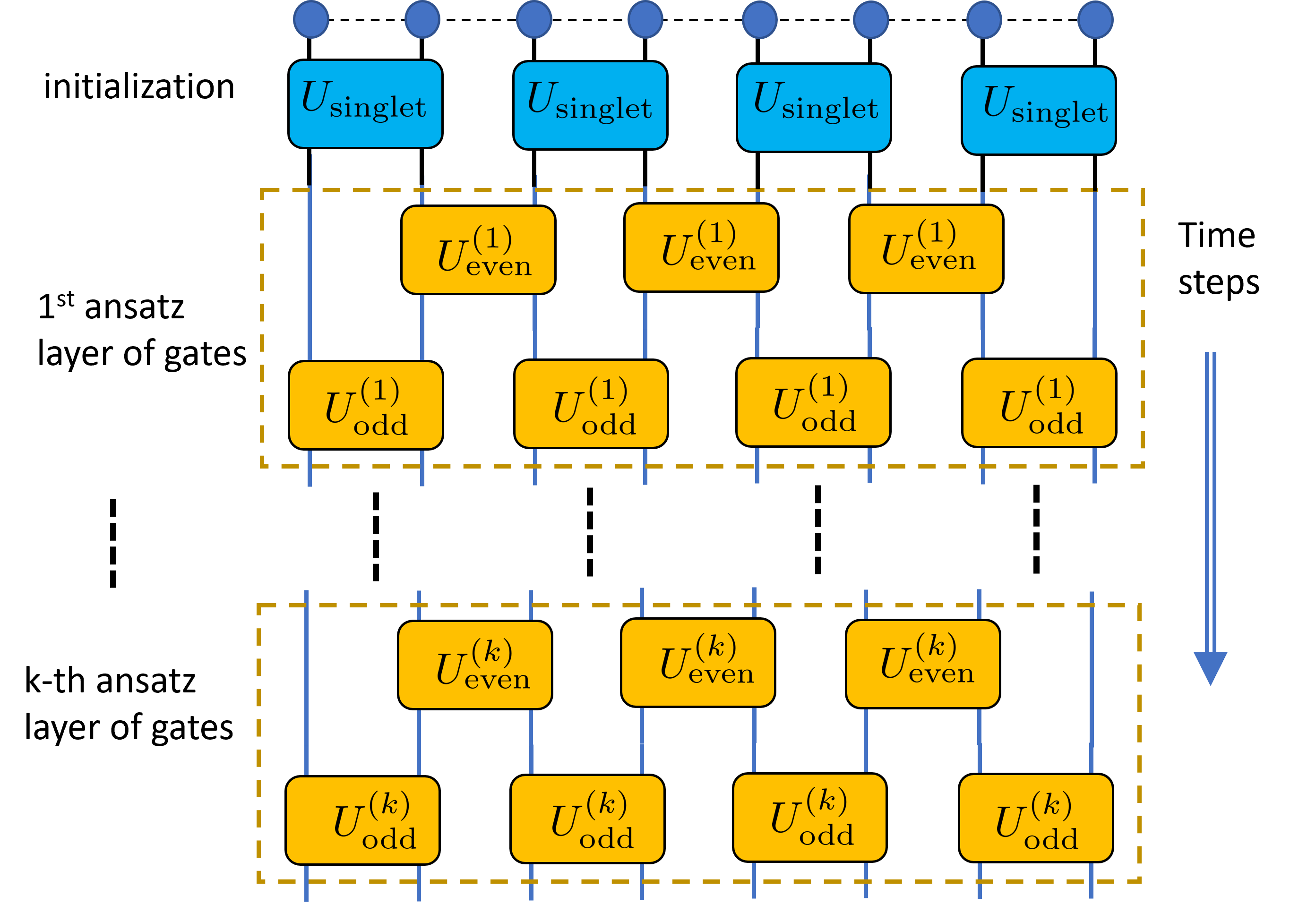} & \includegraphics[width=0.48\textwidth]{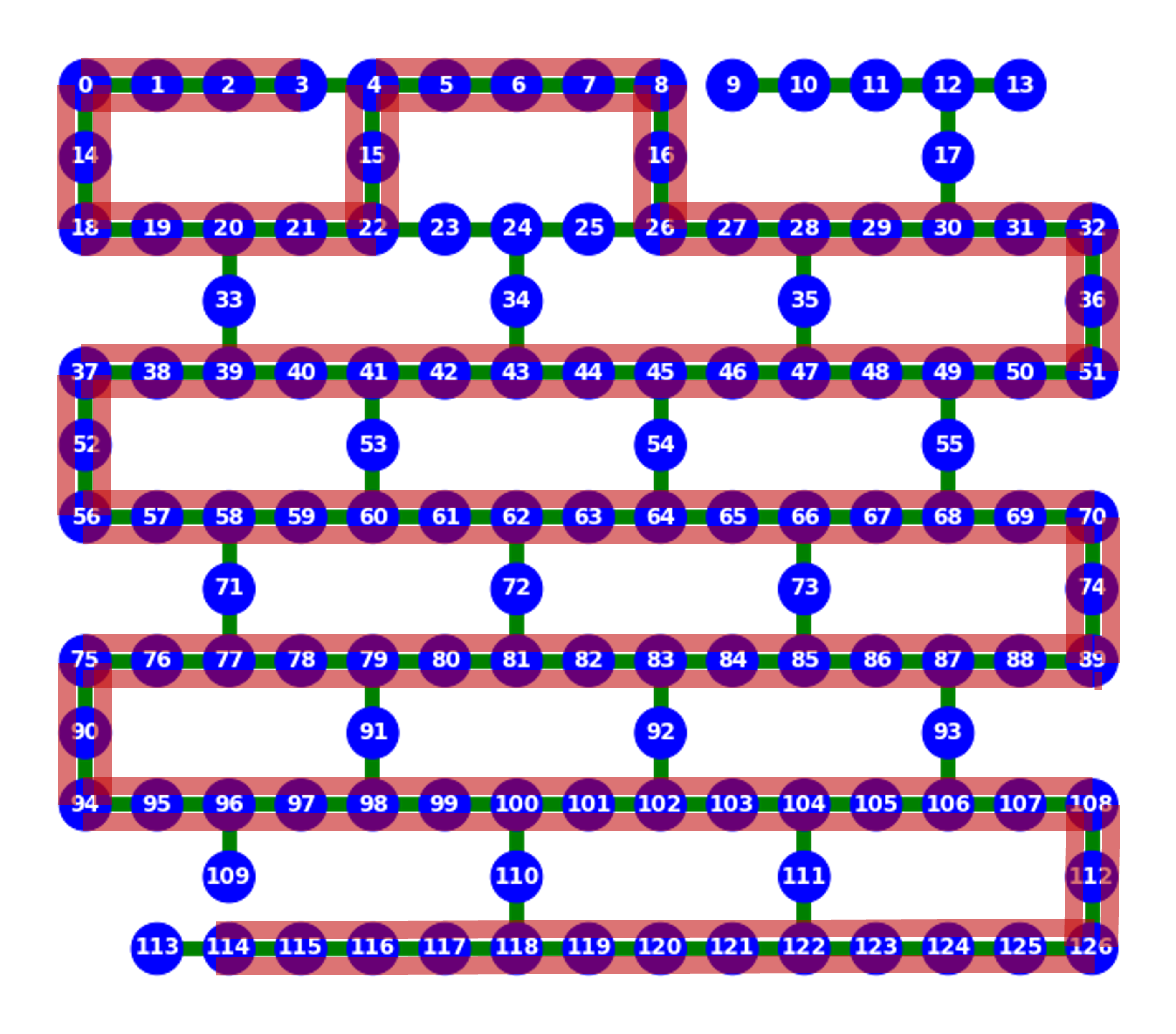}\\
\end{tabular}
\caption{\label{fig:FidelityLayers}  Variational Ansatz,  layout of a 102-qubit quantum computer, and simulation results of the Ansatz.  (a) The variational Ansatz structure, (b) the layout of the 127-qubit {\tt ibm\_washington} backend,  where a chain of 102 qubits is illustrated by the thick, shaded line. 
 }
\end{figure*}

The notion of quantum computers traces back to the works of Benioff~\cite{benioff1980}, Mannin~\cite{mannin1980}, and Feynman~\cite{feynman1982}. In particular, Feynman suggested using quantum computers to simulate other quantum systems instead of using classical computers, giving rise to the notion of a universal quantum simulator~\cite{lloyd1996}. A critical breakthrough was made by Shor, whose quantum factoring algorithm outperforms classical algorithms almost exponentially faster~\cite{shor1994}.  Experimental progress has come a long way, leading to the burgeoning of quantum devices, with the total qubit number now exceeding one
 hundred in the best cases. However, these devices are still regarded as noisy intermediate-scale quantum (NISQ) processors~\cite{NISQ},
  not yet suitable for full-scale quantum error correction and fault-tolerant quantum computation. On the other hand, there are concurrent efforts to develop error mitigation techniques~\cite{Endo,Temme,znekandala,Dumitrescu,Klco,Urbanek,Giurgica-Tiron,wallman2016,kim2021scalable} and algorithms~\cite{NISQRMP} for NISQ devices to realize their potential quantum advantage~\cite{Preskill2012}.  Recent notable experimental achievements include random quantum circuits of around 50 qubits~\cite{GoogleSupremacy,Zuchongzhi}, boson sampling with large numbers of photons~\cite{Jiuzhang,madsen2022quantum},  Hartree-Fock method implementation for quantum chemistry with 12 qubits~\cite{GoogleHartreeFock},  realization of the toric-code state with 31 qubits~\cite{GoogleTopological}, and  quantum walks on a 62-qubit processor~\cite{PanQW}. 
 
  Despite all this  effort and accomplishment, the central question remains of whether NISQ computers can be of practical use for the simulation of large quantum systems and to extract accurate observable values, such as the energy of the simulated quantum states. So far, most experiments with quantitatively accurate results have been limited to small numbers of qubits, around ten or below~\cite{znekandala,Dumitrescu,Klco,Urbanek,GoogleHartreeFock}, with a few others reaching beyond twenty~\cite{kim2021scalable,GoogleTopological}. None of them has demonstrated accurate results over a wide range of  system sizes with the same model and across different devices. There are also challenges to overcome for large-scale experiments (around or over one hundred qubits) with useful outcomes, including the need for   high-fidelity gates and readout as well as  scalable and efficacious approaches to mitigating the effects of noise and errors on the measured observables.

In this work, using nine distinct cloud quantum computers, we present  realizations  of approximate ground states (GS) of spin chains having nineteen different system sizes, ranging from 4 to 102 qubits. To distinguish our work from experiments performed on in-house devices or customized physical
apparatuses, we shall refer to our use of third-party hardware as `cloud experiments,' as well as to make a distinction  from numerical simulations.  We report the  extracted GS energies, accurate to within a few percent level of error, including the inference of the energy density in the thermodynamic limit from these values.     We emphasize that these cloud experiments are not  equivalent to numerical simulations, as the actual devices have substantial noise and errors and devices' condition can drift over time, and sometimes the same submitted jobs can fail. Nevertheless, cloud-based experiments offer a new paradigm for research and development.
To achieve our accurate results, we have designed  a physics-motivated variational Ansatz, and developed efficient  approaches for measuring energies.  We  have utilized our improved, scalable, mitigation methods to extract accurate GS energy values for  large systems, despite the presence of noise and errors in the gates and the readout. 
The introduction of a reference state in the zero-noise extrapolation (rZNE) substantially improves the accuracy of the results. In addition, we have used our procedure to measure the energies of several Ansatz states that have  randomly chosen parameters, and obtained accurate mitigated energy values. 
 Our work thus establishes a simple--yet substantially improved--quantum variational protocol with mitigation, and paves the way for massive use of large NISQ computers for fundamental physics studies of many-body systems, as well as  for practical applications, 
 including optimization problems.

\section{Heisenberg and XXZ models and the Ansatz for ground states.}
Quantum spin systems, such as the Heisenberg~\cite{Heisenberg} and XXZ models~\cite{Kasteleijn,KorepinBook}, have sparked analytical development and understanding of quantum phases and also  served as a testbed for  numerical techniques.
 The Hamiltonian of the  spin-1/2 XXZ spin chain with the open boundary condition reads
 \begin{eqnarray}
 &&\hat{H}_{\rm XXZ}(\Delta)=\sum_{j=1}^{N-1} \hat{h}_{XXZ}^{[j,j+1]}(\Delta) \\
&& =\sum_{j=1}^{N-1} \Big(\sigma_x^{[j]}\sigma_x^{[j+1]}+
\sigma_y^{[j]}\sigma_y^{[j+1]}+\Delta\sigma_z^{[j]}\sigma_z^{[j+1]}\Big),\nonumber
 \end{eqnarray}
 where
  $\Delta$ represents the anisotropy in the coupling. 
  For $\Delta=1$, the model reduces to the isotropic antiferromagnetic
 Heisenberg chain. The model is known to possess
 three distinct quantum phases: (i) a ferromagnetic phase for $\Delta < -1$, where  classical states, such as $|\!\uparrow\uparrow\dots\rangle$ and $|\!\downarrow\downarrow\dots\rangle$, are
 ground states;
 (ii)  a gapless, critical phase, for
 $-1< \Delta < 1$; and (iii) an antiferromagnetic phase for $\Delta>1$.
  We will mainly focus on the range of $\Delta>-1$ with nontrivial ground states. 
 
 In the following, we explain  how we use adiabatic connection~\cite{QAOA} to arrive at a physics-motivated Ansatz,  schematically shown in Fig.~\ref{fig:FidelityLayers}a,  and justify its validity by considering the gap structure through the adiabatic connection, as illustrated in Fig.~\ref{fig:gapsD}. We also analyze how well the Ansatz performs and how its accuracy is improved by increasing the number of layers.

\subsection{Gap structure of the interpolated Hamiltonian and the Ansatz structure from adiabatic evolution}

\begin{figure*}[t]
\begin{tabular}{ll}
(a) & (b)\\
\includegraphics[width=0.48\textwidth]{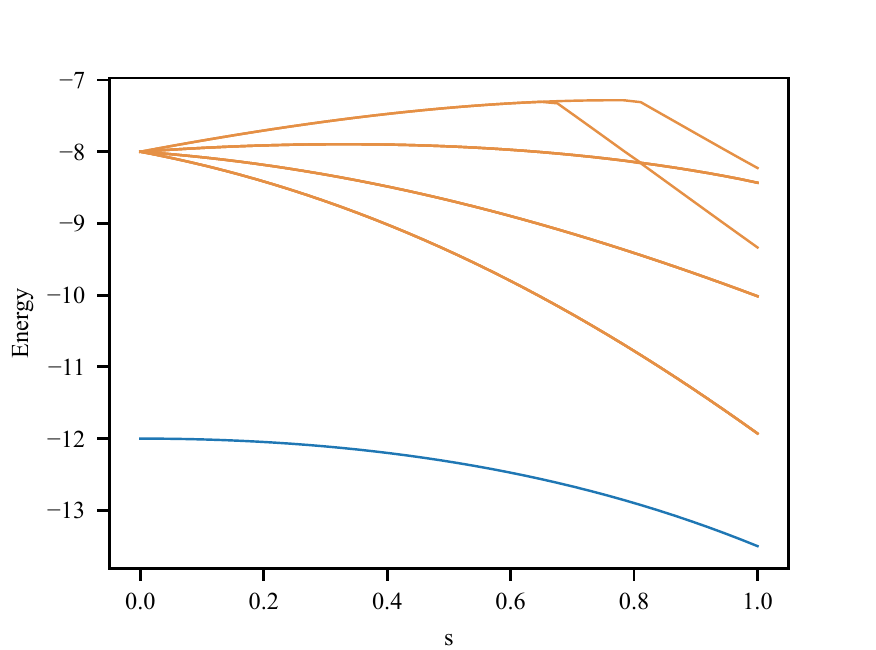}
&
\includegraphics[width=0.48\textwidth]{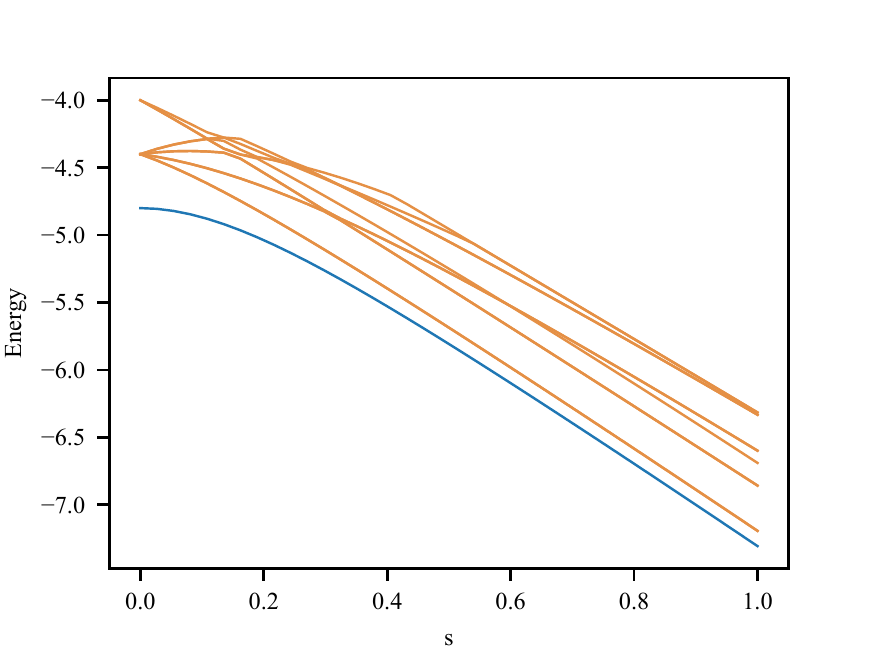}
\end{tabular}
\caption{\label{fig:gapsD} The energy gap of 8-qubit XXZ model with
  open-boundary condition in the Hamiltonian interpolation $\hat{H}(s)$ from one with interaction on odd
  bonds only to one with interaction on all bonds, for (a) $\Delta=1$, i.e. the
  Heisenberg model, and (b) $\Delta=-0.8$.}
\end{figure*}

For the XXZ interaction $\hat{h}_{XXZ}^{[j,j+1]}(\Delta)$ on a bond involving two nearest-neighbor qubits, the singlet pair
$|\Psi^{-}\rangle=(|01\rangle-|10\rangle)/\sqrt{2}$ has an energy value $-2-\Delta$, the triplet
$|\Psi^{+}\rangle=(|01\rangle+|10\rangle)/\sqrt{2}$ has energy $2-\Delta$, and both $|00\rangle$ and $|11\rangle$ (or
equivalently the two entangled triplets $|\Phi^\pm\rangle=(|00\rangle\pm|11\rangle)/\sqrt{2}$) have energy $\Delta$. Note that we have used the notation $|0/1\rangle$ to replace $|\uparrow/\downarrow\rangle$,  the eigenstates of the Pauli Z operator $\sigma_z$.  Thus, the singlet
is the ground state of the simple two-qubit XXZ interaction for $\Delta>-1$.
This means that 
the following Hamiltonian with interaction only on odd numbers of bonds only   even  $N$, 
\begin{eqnarray}
\hat{H}_{\rm odd}&=&\sum_{j=1}^{N/2-1}  \Big(\sigma_x^{[2j-1]}\sigma_x^{[2j]}+
\sigma_y^{[2j-1]}\sigma_y^{[2j]}\nonumber\\
&&+\Delta\sigma_z^{[2j-1]}\sigma_z^{[2j]}\Big),
\end{eqnarray}
has its unique ground state being the product of singlets over
these odd bonds, i.e., a linear chain of valence-bond state,
 \begin{equation}
|\psi_{\rm singlets}\rangle=\frac{1}{\sqrt{2^{N/2}}}\prod_{j=1}^{N/2} (|01\rangle-|10\rangle)_{2j-1,2j}.
\end{equation}
We then expect that
$|\psi_{\rm singlet}\rangle$ is adiabatically connected to the ground state of the XXZ model, by
connecting $\hat{H}_{\rm odd}$ to  the full
XXZ Hamiltonian $\hat{H}_{\rm XXZ}$
via the following linear interpolation,
\begin{equation}
\hat{H}(s)= (1-s) \hat{H}_{\rm odd} + s\, \hat{H}_{\rm XXZ}=\hat{H}_o(s)+\hat{H}_e(s).
\end{equation} 
We regroup it into interaction terms on even and odd bonds, denoted collectively by $\hat{H}_o(s)$ and $\hat{H}_e(s)$, respectively,
and it is straightforward to see that
$\hat{H}_o(s)=\hat{H}_{\rm odd}$, but
\begin{eqnarray}
 \hat{H}_e(s)&=&   s\sum_{j=1}^{N/2-1}  \Big(\sigma_x^{[2j]}\sigma_x^{[2j+1]}+
\sigma_y^{[2j]}\sigma_y^{[2j+1]}\nonumber\\
&&+\sigma_z^{[2j]}\sigma_z^{[2j+1]}\Big),
\end{eqnarray}
is a rescaled version of the XXZ model on even bonds.

We check the spectral properties of this Hamiltonian for small $N$ and
find that $\hat{H}(s)$ is gapped in the range  $s\in[0,1]$ for $\Delta>-1$; see Fig.~\ref{fig:gapsD} for two different $\Delta$ values using 8 qubits.
 This means that the product of
singlets $|\psi_{\rm singlet}\rangle$ is adiabatically connected to the ground state of the XXZ model
via the evolution
  $|\psi(1)\rangle = \mathbf{U}_{\rm evo} |\psi_{\rm singlets}\rangle=e^{-i \int_{s=0}^{1} ds\,\hat{H}(s)}|\psi_{\rm singlets}\rangle$.
Discretizing the evolution operator $\mathbf{U}_{\rm evo}$, we have the following Trotterized approximation
\begin{equation}
\mathbf{U}_{\rm evo}\approx\prod_{l=1}^{N_L} e^{-i \hat{H}(s_l)\delta s}\approx \prod_{l=1}^{N_L} \Big( e^{-i \hat{H}_{\rm e} (s_l)\delta s}e^{-i \hat{H}_{\rm o} (s_l)\delta s}\Big),
\end{equation}
where $N_L$ is the number of discretized time steps or layers and $\delta s=1/N_L$ is the dimensionless  step size. To allow for flexibility, we turn the discretized evolution into a variational form and arrive at the
structure of the Ansatz shown in Fig.~\ref{fig:FidelityLayers}a with gates in the $l$-th layer being
\begin{eqnarray}
  &&  U_{{\rm even/odd}}^{(l)}(\{\theta\})=\mathop{\bigotimes}_{j\in {\rm even/odd}} \\
 && \!\!\!\!\!\!  \big[
e^{-i\theta_{e/o,x}^{(l)}\, \sigma_x^{[j]}\sigma_x^{[j+1]}-i\theta_{e/o,y}^{(l)}\,\sigma_y^{[j]}\sigma_y^{[j+1]}-i\theta_{e/o,z}^{(l)}\, \sigma_z^{[j]}\sigma_z^{[j+1]}}\big], \nonumber
\end{eqnarray}
where $\{\theta\}'s$ are a set of variational parameters,  the subscripts $e/o$ denote the association with  even and odd
bonds, respectively.
Therefore, we arrive at the following $N_L$-layer variational Ansatz state
\begin{eqnarray}
&&  |\psi_{\rm ansatz}(\{\theta\})\rangle\\
  &&=\mathop{\bigotimes}_{l=1}^{N_L}\big[ U^{(l)}_{\rm even}(\{\theta_e\})U^{(l)}_{\rm odd}(\{\theta_o\})\big] |\psi_{\rm singlets}\rangle.\nonumber
\end{eqnarray}
We remark that our construction from the adiabatic connection is similar to how the Quantum Approximate Optimization Algorithm (QAOA) Ansatz originates from discretizing
the Ising interaction and the transverse
field~\cite{QAOA}. 
But they differ in the goal: the QAOA aims to minimize the energy of a
classical Ising Hamiltonian using the transverse-field part as a driver, whereas our goal is to optimize the energy of a quantum Hamiltonian as a whole.  We note that a similar Ansatz for the Heisenberg mode on the kagome lattice  was also studied numerically in Ref.~\cite{kattemolle2021variational}.

\subsection{Creation of singlets}
Each singlet pair in $|\psi_{\rm singlets}\rangle$  can be created from $|00\rangle$ by simple single-qubit gates (the Hadamard $H$ and the Pauli $X$ gates) followed by a CNOT gate,
\begin{eqnarray}
&&|\Psi^-\rangle=\frac{1}{\sqrt{2}}(|01\rangle-|10\rangle)= U_{\rm singlet} |00\rangle   \nonumber\\
&&\begin{array}{c}
\\
=\quad
\end{array}
 \Qcircuit @C=1em @R=.7em {
 |0\rangle&  & \qw& \gate{X} &\gate{H} & \ctrl{1} &\qw  \\
 |0\rangle &  & \qw& \gate{X} & \qw & \targ &\qw                     }       \end{eqnarray}
and, thus, the product of such singlet pairs can be created in parallel with these circuits,
\begin{equation}
\label{eqn:singlets}
 |\psi_{\rm singlets} \rangle =U_{\rm init}|0\dots 0\rangle= \bigotimes_{i=1}^{N/2} U_{\rm singlet}^{[i]} |0\dots 0\rangle, 
\end{equation}
with the superscript $i$ denoting the pair of qubits for the singlet creation. We note that the reverse of the latter part corresponds to Bell measurement,
\begin{equation}
    \Qcircuit @C=1em @R=.7em {
  & \qw&\ctrl{1} & \gate{H} & \qw  \\
 & \qw & \targ &\qw    &\qw            }
\end{equation}
which can be used to measure the energy contribution of a pair of qubits; see below.
 \subsection{Gate decomposition}

\begin{table*}[ht!]
\begin{tabular}{|l|l|l|l|l|l|l|l|l|}
\hline
 $N$ & $\theta_{\rm even}^*$ & $\theta_{\rm odd}^*$ & $E_{\rm ansatz}^*$& $E_{\rm gs}$ & $\epsilon$ & $f$ &  $E_{\rm exp}$ & error \\ \hline
4 & 0.151748 & 0.215765
&-6.464102 &-6.464102 & 0 &1.0000 &-6.5(1.6) & 0.56\%\\ 
6 &  0.141671 & 0.216088
&-9.880996 &-9.974309&0.94\%&0.9923 &-9.9(1.9)$^*$ &0.19\%\\
8& 0.138569 & 0.216093  &
-13.299823&-13.499730&1.48\%&0.9796 & -13.2(2.2) & 2.22\% \\
10& 0.13710 &  0.216102
&-16.719307 &-17.032141&1.84\%&0.9639 &-16.7(1.3)$^*$ &1.95\%\\
12&  0.136248 & 0.216110
&-20.139037 &-20.568363& 2.09\%&0.9462 & -20.3(2.1) & 1.30\% \\
14&  0.135688 & 0.216115 &
-23.558885 & -24.106899& 2.27\% &0.9271 & -23.6(1.8) & 2.10\%\\
16&  0.135293 & 0.216120 &
-26.978800 &-27.646949&2.42\%&0.9072 & -25.8(1.6)$^*$ & 6.68\%\\
18&0.134999 & 0.216123 &
-30.398756 &-31.188044&2.53\%&0.8867 &-30.7(0.7)$^*$  &1.56\%\\
20& 0.134773 & 0.216126 &
-33.818738 &-34.729893&2.62\%&0.8659 & -33.0(0.5)$^*$ & 4.98\%\\
30&  0.134132 & 0.216134&
-50.918850&-52.445423&2.91\%&0.7614 & -50.2(2.0)$^*$& 4.28\%\\
40&  0.133832 & 0.216139 &
-68.019098 &-70.165893&3.06\%&0.6629 & -68.5(2.0)$^*$ &2.34\%\\ 
50& 0.133658 & 0.216141 &
-85.119397 &-87.888441&3.15\%&0.5737 & -85.0(2.8)$^*$ &3.29\%\\
60 & 0.133544 & 0.216143 & -102.219721 & -105.612060 &3.21\% & 0.4946 & -99(4) & 6.26\%\\
70  & 0.133464 & 0.216144 & -119.320058 & -123.336305 & 3.26\% &0.4253 &-125(7) & 1.35\% \\
80 & 0.133405 & 0.216145 & -136.420403 & -141.060947 & 3.29\% & 0.3649& -138.5(2.5)& 1.82\% \\
90 &   0.133359 & 0.216146 &  -153.520754 & -158.785857 & 3.32\% & 0.3126 &-153(5)& 3.64\%\\
98 &  0.133329 & 0.216146 & -167.201038 & -172.965924  & 3.33\% & 0.2760 & -168.1(2.6)& 2.81\%\\
100& 0.133323 &  0.216146 &
-170.621109&-176.510957&3.34\%&0.2675& -173(9) &1.99\%\\
102 & 
 0.133316 & 0.216146  & -174.041180 & -180.055995 & 3.34\% & 0.2592 & -177.5(2.7) & 1.42\%\\
\hline
\end{tabular}
\caption{\label{tb:Heisenberg}  Results related to the open-chain Heisenberg model.
The numerical calculation was done with the MPS method using a bond dimension $\chi=64$.
The `error' in the last column represents the relative error between the experimentally estimated value $E_{\rm exp}$ and the exact ground-state energy $E_{\rm gs}$.  $^*$Note that these values were obtained by averaging results over different backends and/or different groups of physical qubits; see Tables S.2 and S.3     
   for the complete list of results. 
   }
\end{table*}
Let us define the essential two-qubit Rxyz gate that we need,
\begin{eqnarray}
&& R_{xyz}(\theta_x,\theta_y,\theta_z) \\
&& \equiv e^{-i(\theta_x/2) \sigma_x\otimes\sigma_x-i(\theta_y/2)\sigma_y\otimes\sigma_y-i(\theta_z/2) \sigma_z\otimes\sigma_z}, \nonumber
\end{eqnarray}
where a factor of 1/2 is inserted in the definition of the Rxyz gate to match the convention of single-qubit rotation and we have used the tensor product notation `$\otimes$' to emphasize the two-qubit structure in the gate.
We present a decomposition of the Rxyz gate that has a minimum number of CNOTs~\cite{VatanWilliams} (which is three) in the decomposition,
\begin{widetext}
\begin{eqnarray}
\begin{array}{c}
\\
\\
\!\!\!\!\!\!\!\!\!\!R_{xyz}(\theta_x,\theta_y,\theta_z)= \\
\end{array}&&
\Qcircuit @C=1em @R=.7em {
 &\targ & \gate{R_z(\theta_z)}&\qw& \targ &  \gate{R_z(-\theta_y)} &\gate{H} &\ctrl{1}& \gate{S} & \gate{H}&\qw \\
&\ctrl{-1} & \gate{H}\qw & \gate{R_z\big(\theta_x+\frac{\pi}{2}\big)} & \ctrl{-1}  & \qw & \qw & \targ&\gate{S^\dagger}& \gate{H} & \qw
} \nonumber\\
\begin{array}{c}
\\
\\
= \\
\end{array}&&
\Qcircuit @C=1em @R=.7em {
 &\targ & \gate{R_z(\theta_z)}&\qw& \targ &  \gate{R_z(-\theta_y)}  &\targ& \gate{R_x\big(\frac{\pi}{2}\big)} &\qw \\
&\ctrl{-1} & \gate{H}\qw & \gate{R_z\big(\theta_x+\frac{\pi}{2}\big)} & \ctrl{-1}  & \gate{H} &  \ctrl{-1}&\gate{R_x\big(\frac{-\pi}{2}\big)}  & \qw
}
\end{eqnarray}
\end{widetext}
where $H$ is the Hadamard gate,
$R_\alpha(\theta)=e^{-i\theta \sigma_\alpha /2}$ is the single-qubit rotation around $\alpha$-axis ($\alpha=x,y,z$) by
an angle $\theta$, and $S$ is the one-qubit phase gate $S=e^{i\pi/4} R_z(\pi/2)$. 
The gate
$U_{\rm even/odd}(\theta)= R_{xyz}(2\theta,2\theta,2\theta)$ will be used for the Heisenberg model, and for the XXZ model, due to the ZZ anisotropy, we will allow $\theta_z=2\theta_2$ parameter to be independent from $\theta_x=\theta_y=2\theta_1$, and thus $R_{xyz}(2\theta_1,2\theta_1,2\theta_2)$ is needed.

Note that as the circuit action is symmetric with respect to swapping the two qubits, one can flip the circuit in the last line to fit the desired or natural direction of the CNOT gate. One can also replace the Hadamard gate $H$ by a combination of the square root of $X$ gate (or equivalently $R_x(\pi/2)$, which is among the native gates in IBM Quantum Computers), and the phase gate $S$ via the identity
$H=S R_x(\pi/2) S$. Note that $S$ is equivalent to $R_z(\pi/2)$ up to an irrelevant global phase factor,   and, therefore, the circuit can be expressed entirely in terms of IBM Q's native gate set: \{$R_z$, $S_X$, CNOT, $X$\}, where $S_X$ is the square root of $X$.

\begin{figure*}[ht!]
\begin{tabular}{ll}
  (a)  & (b)  \\ 
  \includegraphics[width=0.48\textwidth]{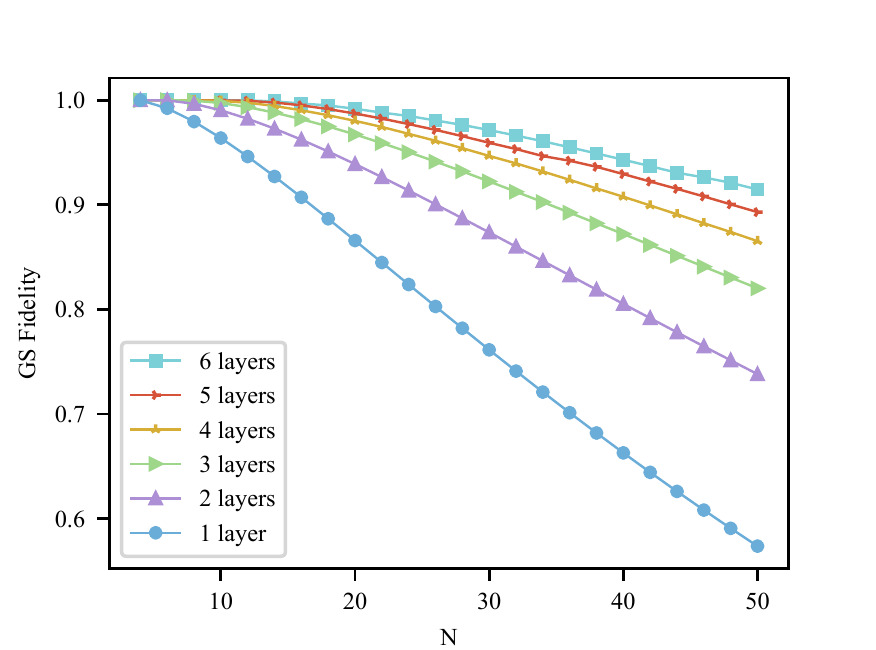} &  \includegraphics[width=0.48\textwidth]{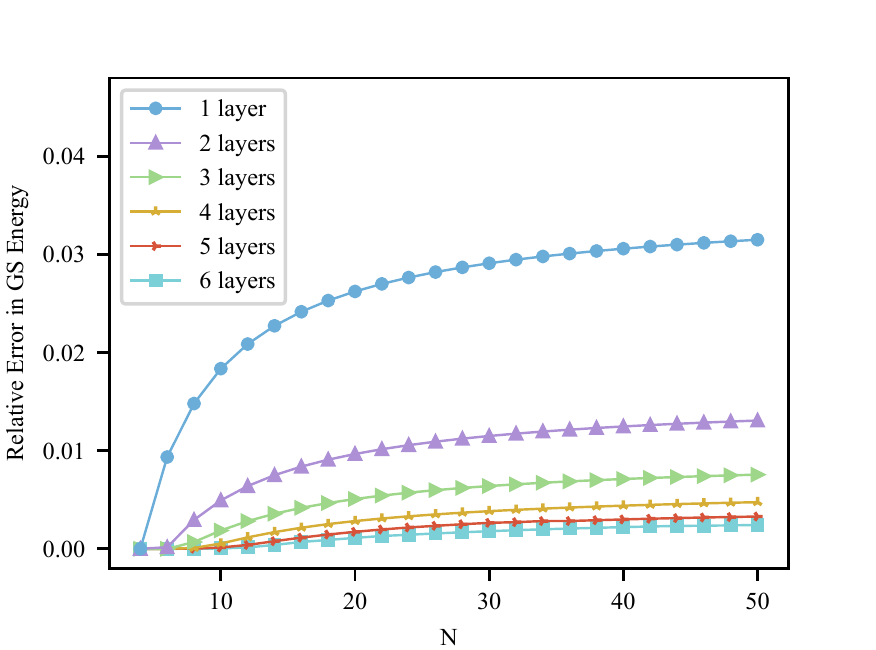}\\
  (c) & (d)\\
\includegraphics[width=0.48\textwidth]{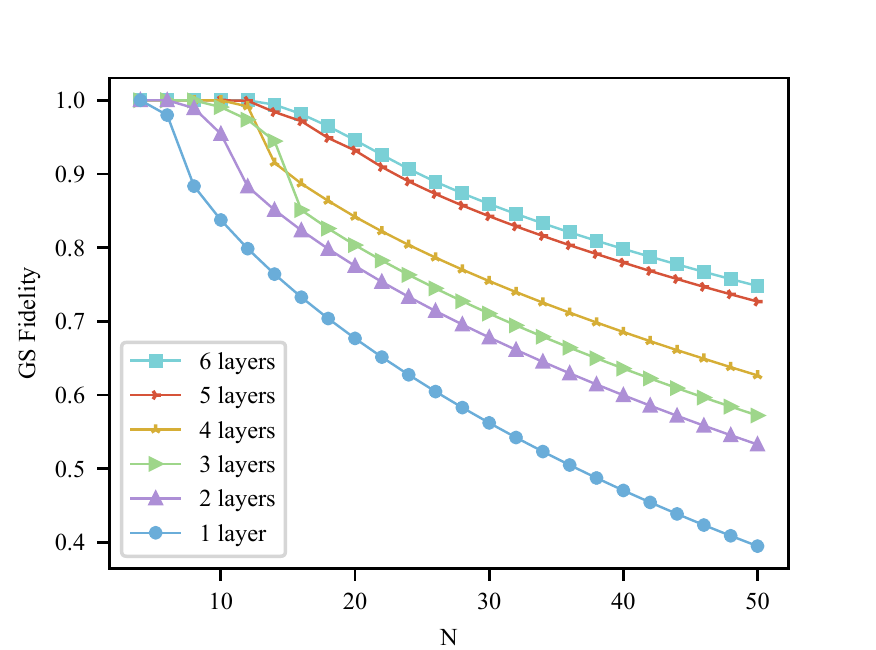} 
&
\includegraphics[width=0.48\textwidth]{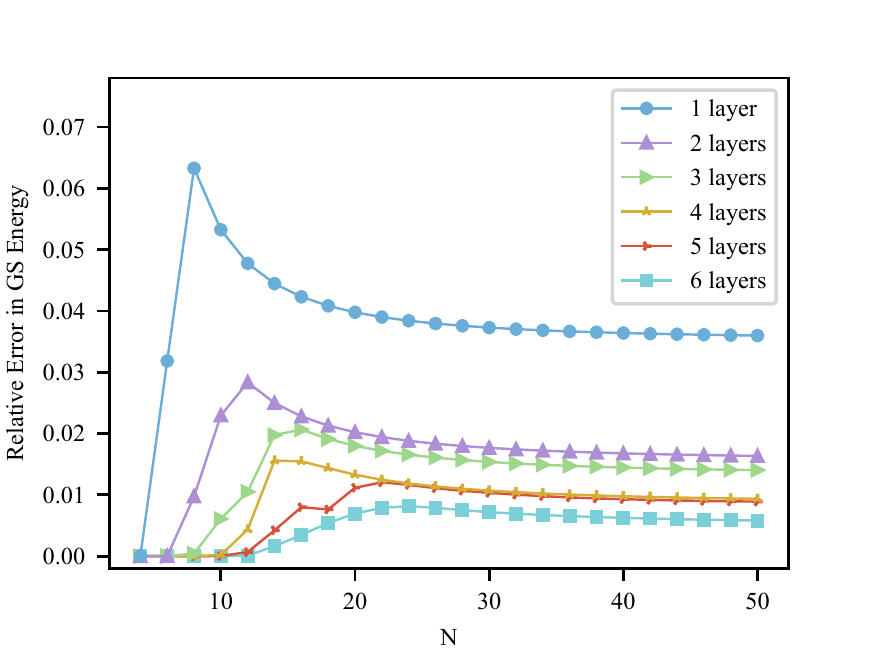}
\end{tabular}
\caption{\label{fig:GSfidelity}  Simulation results of the Ansatz. (a)  the fidelity of the optimal anatz state with the ground state  of the open-chain Heisenberg model vs.   the total number of qubits $N$ for one to six layers in the Ansatz; (b)  the corresponding relative error in the GS energy. 
(c) The fidelity and (d)
  relative energy error  of the optimal Ansatz state and the exact
  ground state of the periodic-boundary-condition Heisenberg model with the
  total number of qubits $N$ for one to six layers in the Ansatz. 
  }
\end{figure*}

\section{Analysis of the Ansatz}
 We have explained how we arrive at a physics-motivated Ansatz,
 \begin{eqnarray}
\label{eqn:U_ansatz}
  &&|\psi_{\rm ansatz}(\{\theta\})\rangle=\mathbf{U}(\{\theta\}) |0\dots0\rangle \\
  &&=\mathop{\bigotimes}_{l=1}^{N_L}\big[ U^{(l)}_{\rm even}(\{\theta_e\})U^{(l)}_{\rm odd}(\{\theta_o\})\big] U_{\rm init}|0\dots0\rangle,\nonumber
\end{eqnarray}
 schematically shown in Fig.~\ref{fig:FidelityLayers}a, 
where  we have used  $U_{\rm init}$ to denote the initialization step that  takes $|0\dots0\rangle \equiv|\!\uparrow\dots\uparrow\rangle$ to a product of singlets  or Bell pairs $|\psi_{\rm singlets}\rangle$ in Eq.~(\ref{eqn:singlets}).


Variational Ansatzes and trial wavefunctions are commonly used in physics. Well-known examples include the BCS wave function for superconductivity~\cite{BCS} and the
 Laughlin wave function for the fractional quantum Hall effect~\cite{Laughlin}.  Despite  not exactly representing the GS, they capture the essential physical properties.
  To analyze how well our  Ansatzes  simulate the GS wave functions and their energy, we  minimize $E_{\rm ansatz}(\{\theta\})$ to obtain the optimal parameters $\{\theta^*\}$ and then compute the GS fidelity~\cite{NielsenChuang00} $f\equiv|\langle \psi_{\rm gs}|\psi_{\rm ansatz}(\{\theta^*\})\rangle|$ and the  error in the GS energy $\epsilon\equiv |E_{\rm ansatz}(\{\theta^*\})-E_{\rm gs}|/|E_{\rm gs}|$, where $| \psi_{\rm gs}\rangle$ denotes the GS  and $E_{\rm gs}$ its exact energy. 
In particular, we find that for $N=4$ and $N=6$, one- and two-layer Ansatzes already can reach  exact ground states for a wide range of $\Delta$ and three-layer Ansatzes achieve the exactness for all $\Delta>-1$~\cite{Joris}.
That variational Ansatzes contain
the exact ground states  is a desired
feature, as it can ascertain optimality of variational parameters. 

In
Table~\ref{tb:Heisenberg},  we show the optimized parameters, energy, and overlap
with MPS diagonalized ground-state wave function using one layer of our Ansatz for the Heisenberg model.
We check that the results agree with the exact computation for the qubit number $N\le 12$.
  As expected, the fidelity decreases with the number of qubits, but much slower than exponentially.
In contrast, the approximate ground-state energy  seems to reach about 3\% of error even
for large chains using just one layer in the Ansatz (e.g. 3.33\% even for
$N=100$).

Using one to six layers in the Ansatz, we compare the fidelity and energy error vs.  $N$ in Fig.~\ref{fig:GSfidelity}ab for the open chain.   For large systems, we use matrix product states (MPS)~\cite{PEPS,MPS1,MPS2,DMRG} for these calculations. The results improve substantially with increasing layers: with six layers and $N=50$,  the fidelity is above 0.9 
and the accuracy in the GS energy is above 99.75\%. 
For  the periodic chain, given our Ansatz breaks the translation
invariance (down to two sites), it will take a few layers to approximately restore the invariance. Overall, we do see general improvement in both quantities as the number of layers in the Ansatz increases. 


\section{Cloud experiments, rZNE and results.}
We have performed cloud experiments by creating the (one-layer) Ansatz states and measuring their energies on nine different backends of IBM Q, which contain 27, 65 and 127 qubits on three types of layouts (see Fig.~\ref{fig:Backends}
  and Table~\ref{tab:backends} in the Appendix).
The cloud experimental results of the 19 different sizes (ranging from 4 to 102 qubits) of Heisenberg chains  are also summarized in Table~\ref{tb:Heisenberg}; their relative errors with the ground-state energy values are within a few percentages.  We will explain below how the experimental results and their mitigated values were obtained. 
For the purpose of demonstration, we mostly use the numerically optimized parameters to run the state creation circuits. But still, we also test the feasibility of the hybrid quantum-classical approach by performing cloud experiments  with random parameters below. 

We first discuss different approaches to measure the total energy and then the readout-error and gate-error mitigation methods to extract estimated values from experiments. In particular, we will introduce a reference state in doing the gate-error mitigation.
\subsection{Measuring energy: several approaches}
\label{sec:meas_ob}
Here we describe three approaches that we have used to measure the energy expectation value. Ideally the three approaches should give the same results and we have indeed tested all three  experimentally and verified that they give the same results within a few percentages of errors for small systems; see Fig.~\ref{fig:xxzResult} in the Appendix.

\smallskip\noindent (1) \textbf{Tomography-based approach}. For the models we consider, the Hamiltonian
terms are of the form $\sigma_{\alpha}^{[j]}\sigma_{\alpha}^{[j+1]}$, where $j$ is the site number
and $\alpha$ is the spin direction (x, y or z). Naively, if we can obtain the reduced
density $\rho_{j,j+1}$ matrix for the pair $(j,j+1)$ then we can calculate the
energy contribution from ${\rm Tr}(\rho_{j,j+1}\sigma_{\alpha}^{[j]}\sigma_{\alpha}^{[j+1]})$. But this 
requires state tomography and seems to need to run $9N_{\rm bond}$ different
circuits for the total energy, where $N_{\rm bond}$ is the number of nearest pairs or bonds, e.g. $N-1$ for an open chain and $N$ for a periodic one. However, we can improve the efficiency by performing
the state tomography in parallel. Doing so, we  just need two sets of state
tomography circuits (for even and odd bonds respectively) to obtain the reduced density matrices of neighboring pairs of qubits. We will later discuss the measurement mitigation on pairs of qubits associated with bonds in order to extract reliable energy contribution. Doing the tomography in parallel reduces the number of circuits to measure to
$9 \times 2$, which is independent of the model size. The benefit of this is
that any one- and two-qubit observables are readily available, such as the local spin observables and the concurrence which 
quantifies nearest-neighbor entanglement, and it applies to all nearest-neighbor
interacting spins. (Extension to finite-ranged interaction is straightforward
but requires more sets of measurements and multi-qubit state tomography.) For our cloud experimental results on measuring the concurrence in a 8-qubit XXZ chain, see Fig.~\ref{fig:ConcurrenceN8} in the Appendix.

\smallskip\noindent (2) \textbf{XYZ measurement}. The second, slightly reduced
measurement method is to measure separately the two neighboring qubits on each
bond in basis $\alpha$ and then average over the classical results
$\overline{\sigma_{\alpha}^{[k]}\sigma_{\alpha}^{[k+1]}}$, treating $\sigma_\alpha=\pm1$ from the measurement outcome assignment. This naively
requires $3N_{\rm bond}$ different circuits for the total energy.
 But a much simplified implementation is to measure all qubits in $\alpha$ basis and
calculate the average $\overline{\sigma_{\alpha}^{[k]}\sigma_{\alpha}^{[k+1]}}$ for all bonds. This
only requires 3 different measurement settings to obtain the total energy. Such
simplification is applicable for the same reason mentioned in the previous approach (1). 

\smallskip \noindent (3) {\bf Bell measurement}. There is another method that uses Bell-state
measurement on all bonds. It exploits the specific form of $\hat{h}_{XXZ}^{[j,j+1]}(\Delta)$, for which  $|\Psi^{-}\rangle\equiv(|01\rangle-|10\rangle)/\sqrt{2}$ has energy $-2-\Delta$, the
triplet $|\Psi^{+}\rangle\equiv(|01\rangle+|10\rangle)/\sqrt{2}$ has energy $2-\Delta$, and both $|00\rangle$ and
$|11\rangle$ (or equivalently $|\Phi^\pm\rangle\equiv(|00\rangle\pm|11\rangle)/\sqrt{2}$) have energy $\Delta$. We
can identify the Bell state on a particular bond with Bell measurement. The
energy contribution of that bond is the energy corresponding to the Bell state obtained from the measurement outcome.
The total energy can be calculated by adding up every bond's energy contribution.
In practice,  the Bell measurement requires a short circuit including a CNOT gate, and the effect of its error can be mitigated; see below. 

Naively, this approach of measuring energy requires $N_{\rm bond}$
different measurement patterns appended at the end of state creation as readout
for the total energy if each bond is measured separately. However, we can divide
the bonds into even and odd groups, as above, and can perform the Bell measurement in
parallel within each group. Then we only need to perform {\it two}
different sets of measurements. This turns out to be the approach we used to perform large-system (up to $N=102$ qubits) cloud experiments on real devices to obtain the total energy.


 We mainly use  the Bell-state approach to measure the  system's energy, as it requires the least resource compared to two other approaches.

\subsection{Error mitigation}
\noindent {\bf Measurement/Readout Error Mitigation}. For superconducting qubits, the readout error can be as large as 10\% or more and it is therefore crucial to mitigate the 
measurement error in order to calculate the correct energy of the created state on the real device. Due to the  expanding deployment of cloud quantum computers, the interest in the issue of state preparation and readout error has recently been rekindled~\cite{keith2018joint,chen2019detector,geller2021efficient,maciejeswski2020,CTMP,M3Miti}. The key idea is to first characterize the measurement pattern dependent on the state input, such as from the detector tomography or simply measuring the probability matrix ${\cal M}$ that relates the input states to the measured outcomes, i.e. $\vec{P}_{\rm measured}= {\cal M} \vec{P}_{\rm ideal}$, where $\vec{P}_{\rm measured}$ and $\vec{P}_{\rm ideal}$ represent respectively the measured and ideal probability distribution. By properly inverting the relation with the constraint that the outcome distribution $P_{\rm ideal}$ be non-negative, one can obtain the mitigated distribution to evaluate the observables. 

For $N$ qubits, the complete matrix ${\cal M}$ is of size $2^N\times 2^N$ and requires preparation of $2^N$ computational states, thus is not efficient and is only doable for a small number of qubits. 
As the models we consider here contain only nearest-neighbor interactions, we are mainly concerned with measurement mitigation for pairs of qubits in a bond, i.e., involved in the interacting Hamiltonian, and such simplification allows us to deal with large systems in a practical way. We can perform readout mitigation pairwise for the nearest-neighbor two qubits on all bonds. Similar to the energy measurement, this can be reduced to two sets of mitigation, i.e., on pairs of even and odd bonds, respectively. Each mitigation requires 4 different inputs from all two-qubit computational basis and measurement in the same basis gives rise to a $4\times4$  matrix ${\cal M}$, which we can then use to infer the ideal two-qubit measurement distribution so as to obtain the mitigated energy contribution.

\smallskip \noindent {\bf Bell-measurement Mitigation}. In our cloud experiments with large numbers of qubits, the local energy for a pair of qubits is obtained by measuring in the Bell-state basis, which uses an inverse circuit for Bell-state preparation and involves CNOT gates.  To mitigate potential errors caused by imperfect CNOT gates, we adopt the above readout mitigation for the Bell measurement. Specifically, for each pair in the bonds, we prepare the four Bell states and then immediately measure qubits pairwise in the Bell-state basis, such as the circuit shown previously, to obtain a $4\times4$ Bell-state assignment matrix ${\cal M}_{\rm Bell}$ for each pair. With this we can mitigate the outcome distribution and hence the energy value obtained from the Bell-state measurement. 

\smallskip \noindent {\bf Gate Error Mitigation}.   By doing readout mitigation we are probing the properties of the state actually created on the quantum devices. However, the observable expectation is affected by gate errors as well that prevent us from obtaining the idealized value. In order to estimate the latter, prior works have considered pulse and gate error mitigation by extrapolating to the zero-error limit~\cite{Endo,Temme,znekandala}, and this is an extrapolation of the physical observables, not the actual observable values associated with the quantum states created. Nevertheless, it is still important to see how well  quantum computers can estimate these values despite the noise and errors, especially in the regime where direct classical calculations might not be feasible.

However,  in order to perform accurate gate mitigation, one needs to have substantial access to   the hardware performing pulse-level optimization and operations~\cite{znekandala}, which is still not practical for dealing with a large number of qubits. (Note that recent experiments have been carried on 26 qubits using pulse-level zero-noise extrapolation~\cite{kim2021scalable}.)
Instead, we will use the gate-level zero-noise extrapolation (ZNE) approach discussed  in Refs.~\cite{Dumitrescu,Klco,Urbanek,Giurgica-Tiron}. In particular, our approach builds on the idea in Ref.~\cite{Giurgica-Tiron} and we prepare the circuits to create $|\psi_n\rangle= \mathbf{U}(\mathbf{U}^{-1}\mathbf{U})^n|0...0\rangle$, where $n$ is a non-negative integer and $\mathbf{U}=\mathbf{U}_{\rm var}(\{\theta\})U_{\rm init}$, as in Eq.~(\ref{eqn:U_ansatz}) of the main text, denotes the circuit to prepare the Ansatz state from the fiducial state $|0\dots0\rangle$, i.e., $|\psi_{\rm ansatz}\rangle=\mathbf{U}(\{\theta\})|0\dots0\rangle$, and then use several forward-backward repetitions in $\mathbf{U}$  to evaluate the observable ${\cal O}_n=\langle\psi_n |\hat{O}|\psi_n\rangle$, as a function of $n$. Ideally,  different $n$ should give the same state and hence the same value for the observable $\hat{O}$. However, noise and errors spoil this and the state with larger $n$ should be noisier. The extrapolation to the gate-level zero-noise limit  is done by a fitting to ${\cal O}_n$ with  $m=2n+1\rightarrow0$ limit.   

    \smallskip \noindent {\bf Reference-state Gate Error Mitigation}. Building on this, we propose to use a reference state (or possibly multiple ones), which is contained in the Ansatz family,  for example, the product of Bell pairs that we use below (via setting all $\theta$'s  to zero, i.e. $|\psi_{\rm singlets}\rangle=\mathbf{U}(\{\theta\}=0)|0\dots0\rangle)$, with a known exact energy value, to improve the extrapolation of the energy value or other observables. Running the energy cloud experiment for this reference state with the above gate-level mitigation, we obtain the naively-extrapolated experimental value and hence the possible mismatch with the exact value. Using such knowledge  for the reference state as a calibration, we can estimate the expected value of the Ansatz state from the naive experimental value.  Combining both gate and readout error mitigation, we are able to reach the accuracy of the extrapolated energy with a few percentages of the exact value for all ranges of the qubit number that we have tested on real devices. We expect that this reference-state ZNE (rZNE)  may be applied to the general VQE platform. It does not require additional circuits from randomized compiling, as done, e.g. in Ref.~\cite{Urbanek,wallman2016}, but averaging the results from these randomized circuits can be used to further improve the accuracy.

\begin{figure*}

\begin{tabular}{ll}
(a) & (b)  \\
\includegraphics[width=0.45\textwidth]{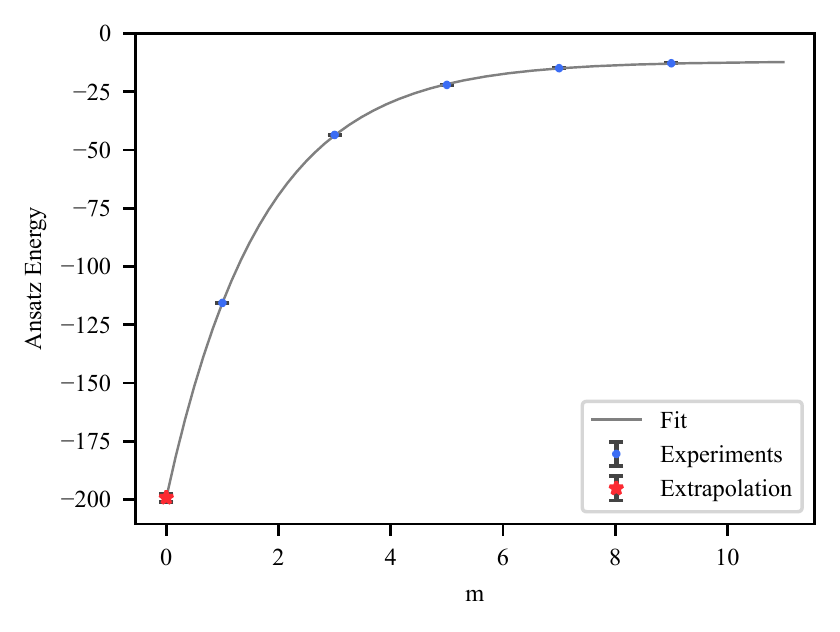}
&
\includegraphics[width=0.45\textwidth]{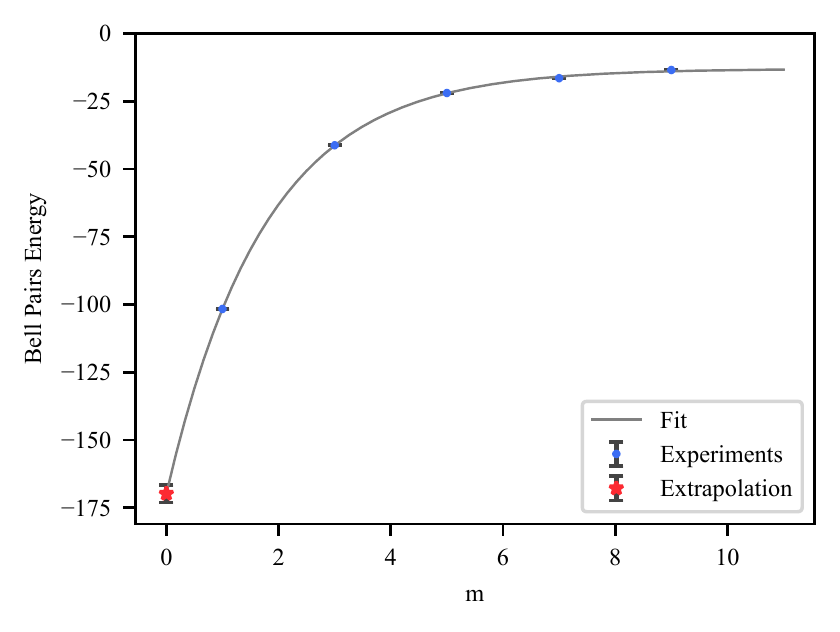}\\
(c) & (d) \\
\includegraphics[width=0.45\textwidth]{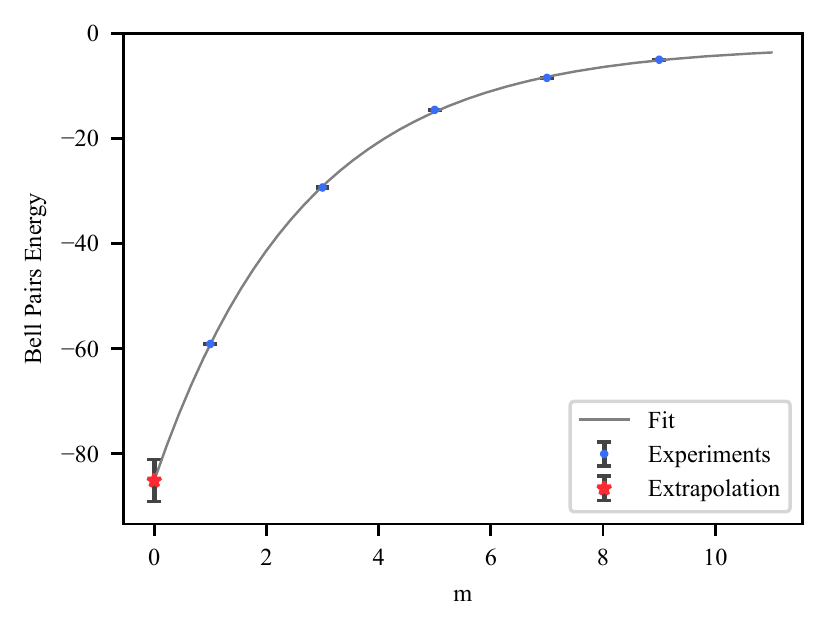}
&
\includegraphics[width=0.45\textwidth]{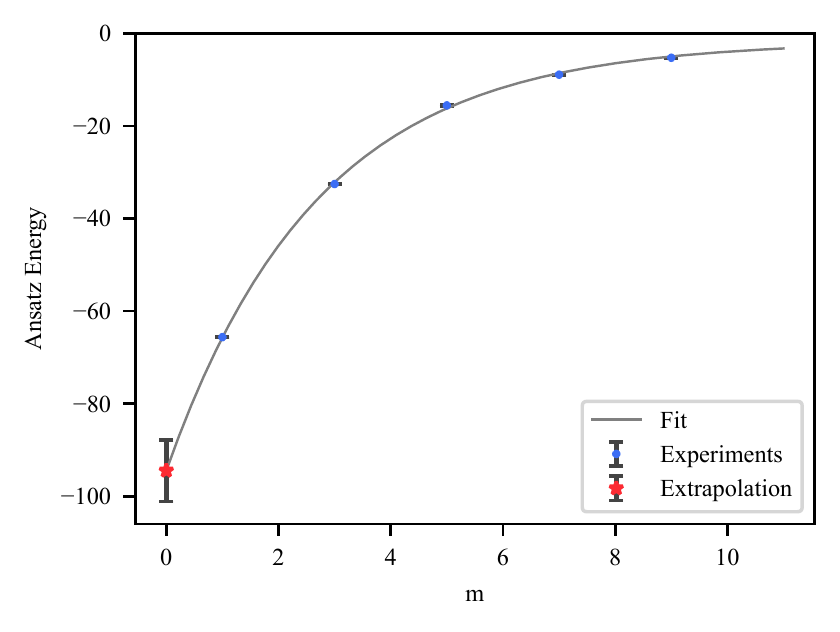}\\
\\ (e) & (f) \\
\includegraphics[width=0.45\textwidth]{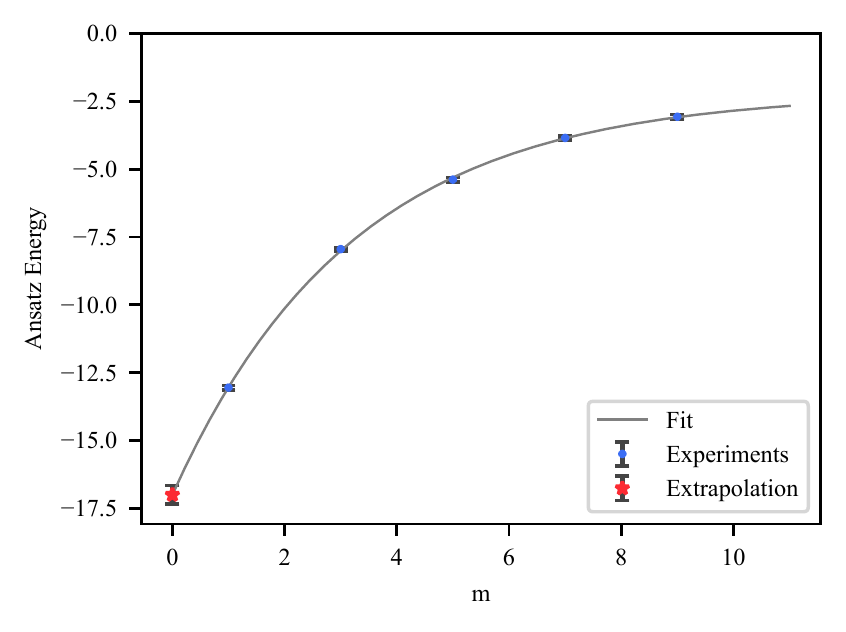}
&
\includegraphics[width=0.45\textwidth]{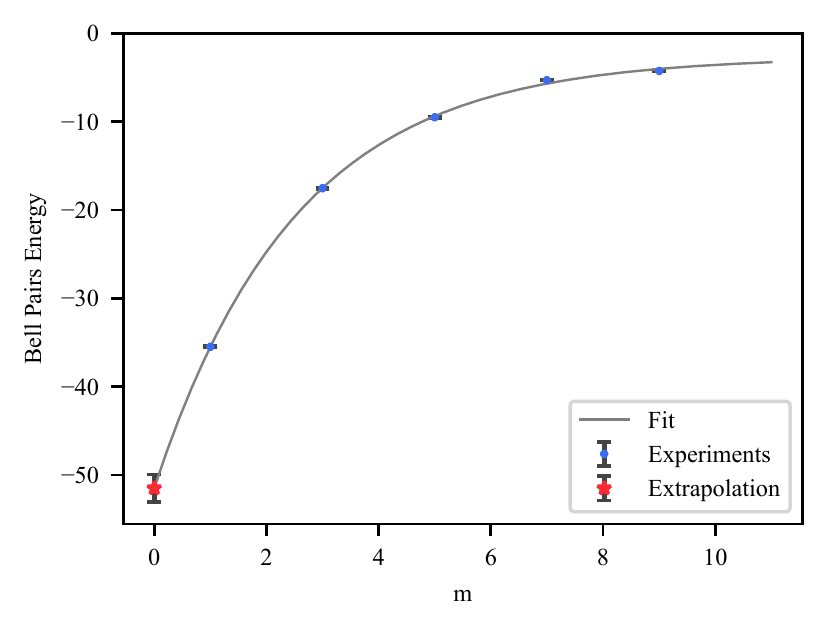}
\end{tabular}
\caption{\label{fig:zne102QB} Zero-noise gate-error extrapolation for cloud experimental realizations of optimal Ansatz states of Heisenberg spin chains, along with the use of Bell pairs. The data points presented were already processed by measurement-error migitation. (a) \& (b) a 102-qubit Heisenberg chain on {\tt ibm\_washington} for the optimal Ansatz state and Bell pairs, respectively; (c) \& (d), similarly for a 50-qubit Heisenberg chain on {\tt ibmq\_brooklyn}; (e) \& (f): results for two 40-qubit random Ansatz states on {\tt ibmq\_brooklyn}. 
In (e), the parameters $[\theta_{\rm even},\theta_{\rm odd}]=[3.5,0.7]$ were used and the exact Ansatz energy is 
  $-16.0669$. 
  In (f),  parameters  $[0.3,1.7]$ were used and the exact Ansatz energy is   -48.0625.
 Separate cloud experiments (results not shown in plots) with 40-qubit Bell pairs gives a naive extrapolation of the Bell pairs energy to be -67.0(4.0), whose ideal value is $-60$. The migtigated values with the reference state for (e) and (f) are $-15.4\pm 0.7$ and $-46.1\pm 2.4$, respectively.
}
\end{figure*}

\begin{figure}
\centering
\includegraphics[width=0.49\textwidth]{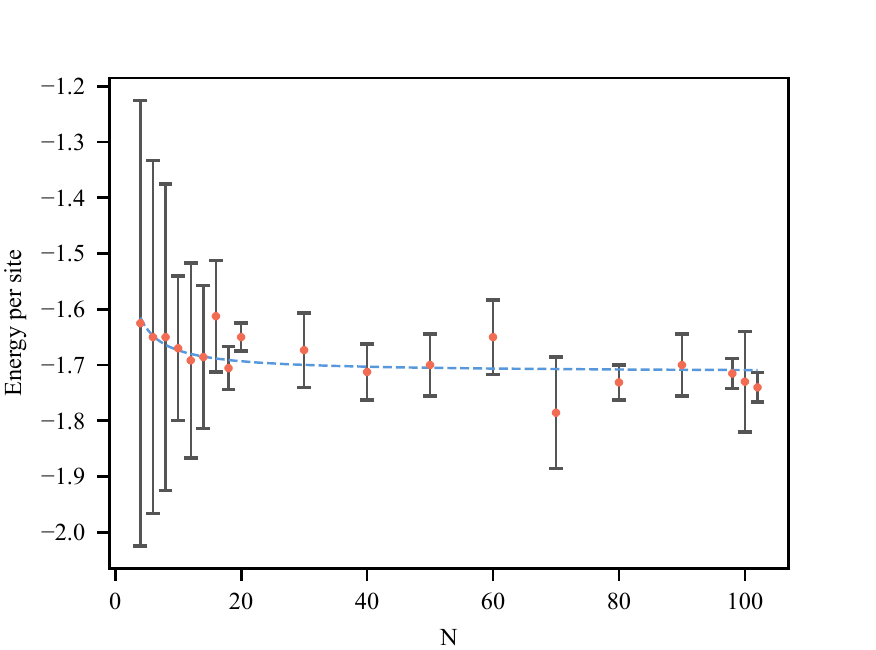}
\caption{\label{fig:energydensity}  Approximated ground-state energy per site vs. the total number $N$ in the spin chain  obtained from cloud experiments performed on various IBM Q backends using the one-layer variational Ansatz. The dashed line is a fit from the data: $-1.713+ 0.393/N$ and the approximated ground-state energy at the $N\rightarrow\infty$ limit is $-1.713\pm 0.046$, which is compared to the exact result from the Bethe Ansatz solution $4(\ln 2 -1)\approx -1.773$.
}
\end{figure}

\subsection{Cloud experimental results: Reference-state zero-noise extrapolation applied}

 \begin{figure*}[ht!]
\centering
\begin{tabular}{ll}
(a) & (b) \\
\includegraphics[width=0.45\textwidth]{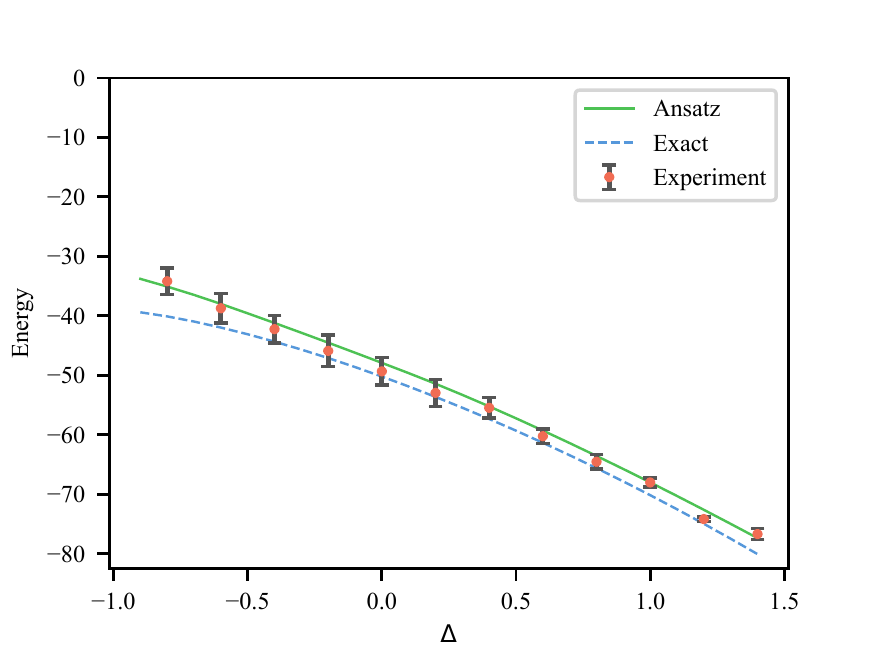} &
\includegraphics[width=0.43\textwidth]{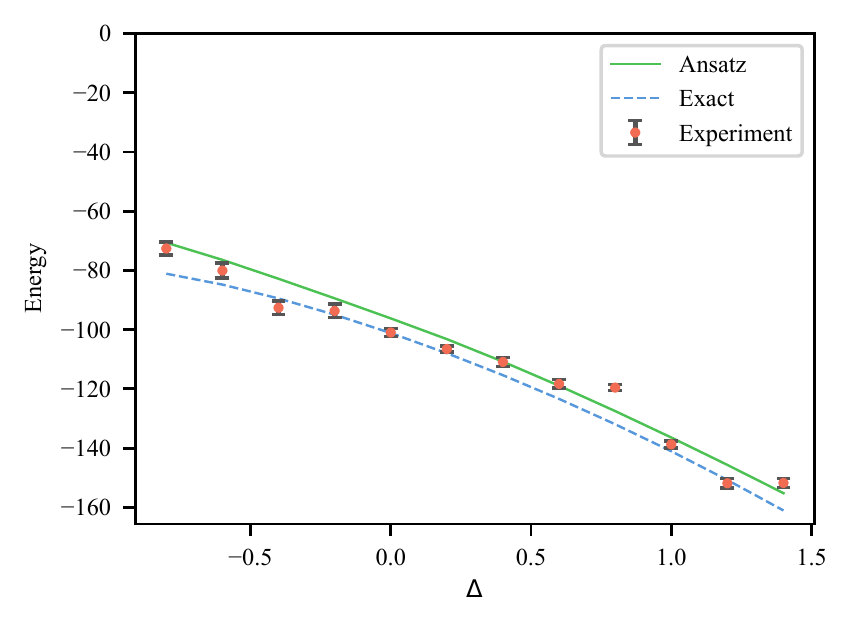}
\end{tabular}
\caption{\label{fig:xxzResultBroolyn}  The mitigated energy results for XXZ spin chains. (a) A 40-qubit XXZ chain on {\tt ibmq\_brooklyn} and (b) an 80-qubit XXZ chain on {\tt ibm\_washington}. The information about which physical qubits were used is listed in the Appendix.
}
\end{figure*}

From the 102-qubit Heisenberg-chain experimental data in Fig.~\ref{fig:zne102QB}ab, we  fit the total energy of the optimal Ansatz state and Bell pairs to  a form  $f_E(m)= a \exp(- b m) +c$ and obtain their respective ZNE values  -199.2 and -169.8.  The  energy of Bell pairs with $N$ qubits is known exactly, $-(2+\Delta)N/2$, which is $-153$ when $N=102$ and $\Delta=1$.  The two values for the Bell pairs enable us to naively correct the Ansatz state energy from $-199.2$ to  
 $-199.2/ (169.8/153)\approx-179$,  close to the  numerical MPS value $-174.04$.  
 
 The noise in real devices is very complex, but a simplified model on neighboring two qubits is a depolarizing channel: $\rho_2\rightarrow (1-p_m) \rho_2 + p_m I\otimes I/4$. One  naively expects that  $p_m=1- e^{-b m}$ and this implies that $f_E(m)=E e^{-b m}$. But in our fitting above, we observe a nonzero residual value $c$.  This means that we should  rescale only  the drop (i.e., $a$) from $m=0$ to $m\rightarrow\infty$. 
 To be more precise, the rescale factor $r$ is obtained via $a_B\cdot r + c_B = E_{\rm bell}$, where $E_{\rm bell}$ is the exact Bell pairs energy and  the subscript $B$ in $a$ and $c$ denotes the parameters obtained from fitting the cloud experiments for Bell pairs. Assume the cloud experiments for the optimal Ansatz experience similar noise and errors, as their circuit structure and depth are identical (except the rotation parameters),   we obtain the extrapolated experimental Ansatz energy to be $E_{\rm exp}=a_E \cdot r + c_E$, where the subscript $E$ denotes the parameters obtained from fitting the cloud experiments for the Ansatz. 
 Applying this to the 102-qubit cloud experiment, we obtain almost the same  result (up to rounding): $E_{\rm exp}=-179.1\pm 3.1$. This is how all the other reported data  were obtained. From our experience, the results obtained this way do not differ much from the naive rescaling in most of our cloud experiments.

 We note that in this set of  cloud experiments, the maximal CNOT depth is 63 and the maxmial total number of CNOTs used is 3186. Similar experimental results on 50 qubits in {\tt ibmq\_brooklyn} are shown in Fig.~\ref{fig:zne102QB}cd. 
  We also note that, in Ref.~\cite{Urbanek}, $|00\dots0\rangle$ was used as a reference state in a circuit that is used to extract the depolarizing rate. Additional randomized instances (e.g. 448 copies) of each main circuit were needed for averaging~\cite{Urbanek}, in addition to tripling and quintupling all CNOT gates for ZNE~\cite{Klco}.

\paragraph{Cloud experimental results for the Heisenberg model on various sizes.} Table~\ref{tb:Heisenberg} has a summary of cloud experimental results of Heisenberg chains  with 19 different sizes (ranging from 4 to 102 qubits) and some were averaged over several different sets of qubits or different devices. We refer to Tables~\ref{tab:127} and~\ref{tab:27and65} 
  for a comprehensive list of 39 mitigated results  on nine different backends.  These  backends possess different qubit numbers, quantum volumes, and noise and error rates (see Table~\ref{tab:backends}), 
  but  the success across all these backends (with varying numbers of qubits used) demonstrates the utility of such a simple and scalable rZNE approach. 
  With these results, we can, for example,  extract the energy per site in the thermodynamic limit (see also Fig.~\ref{fig:energydensity}), 
which yields a value of $-1.713\pm 0.046$ that agrees with the exact Bethe Ansatz calcuation, $4(\ln 2 -1)\approx -1.773$~\cite{KorepinBook}, within 3.4\% of deviation. 


\begin{figure}[h!]
\centering
\includegraphics[width=0.45\textwidth]{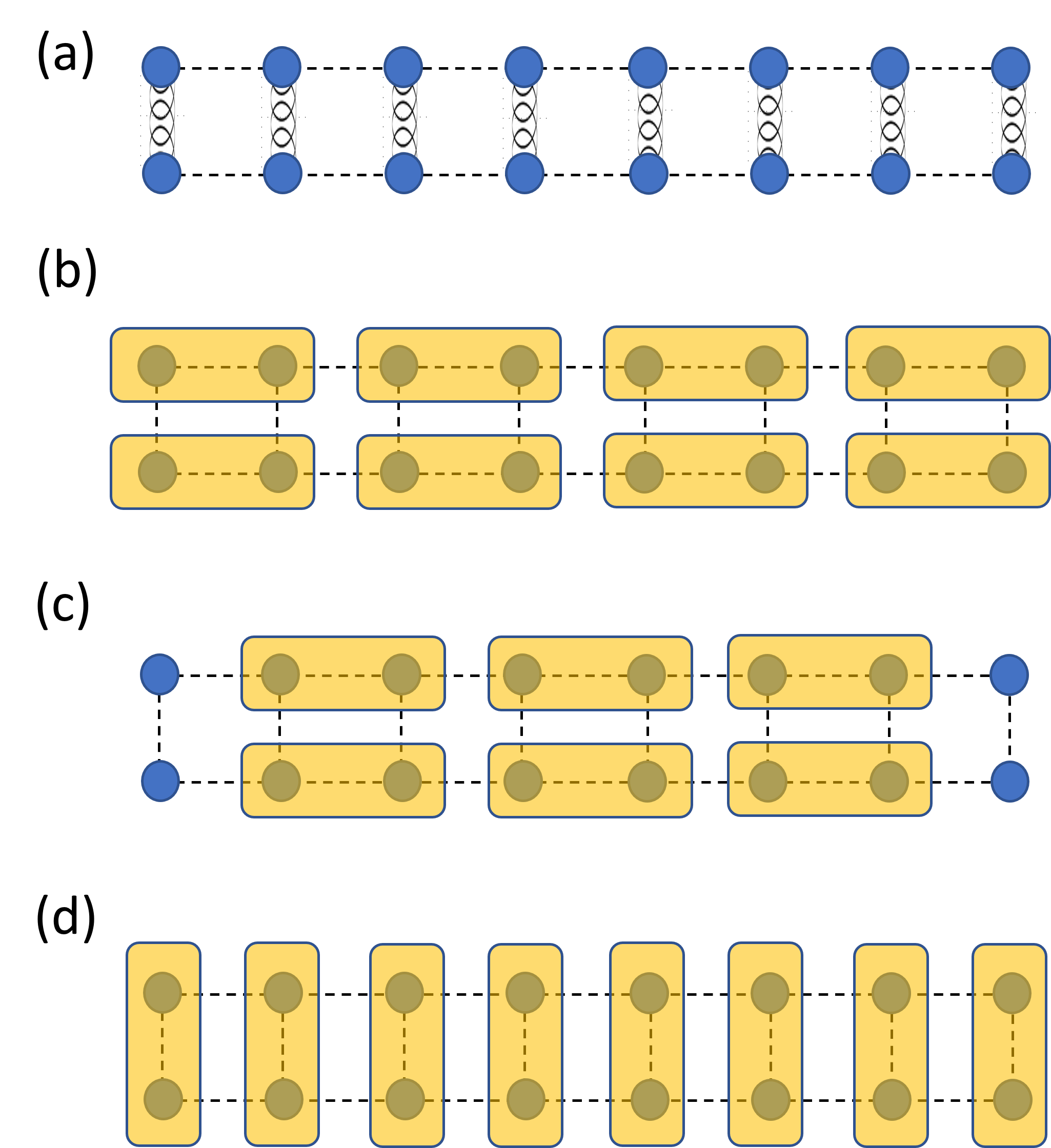}
\caption{\label{fig:LadderAnsatz}  The Anastz for the two-leg ladder. (a) We
  initialize the state in a product of singlets which are formed between the
  upper spins and the lower spins. One layer Ansatz includes (b), (c) and (d),
  where the gates are indicated by shaded rectangles.}
\end{figure}

\begin{figure*}[t!]
\begin{tabular}{ll}
(a) & (b)\\
\includegraphics[width=0.48\textwidth]{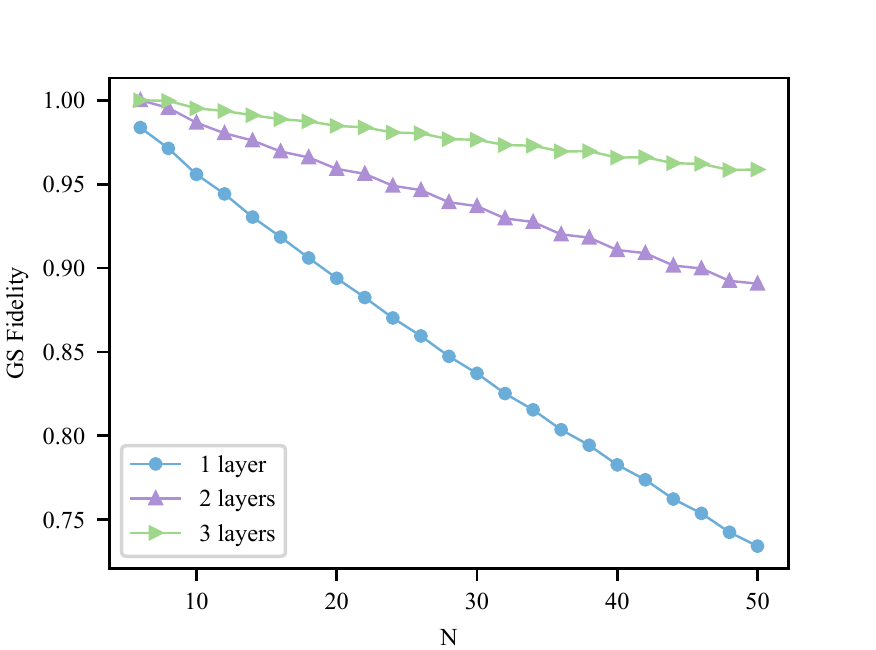}
&
\includegraphics[width=0.48\textwidth]{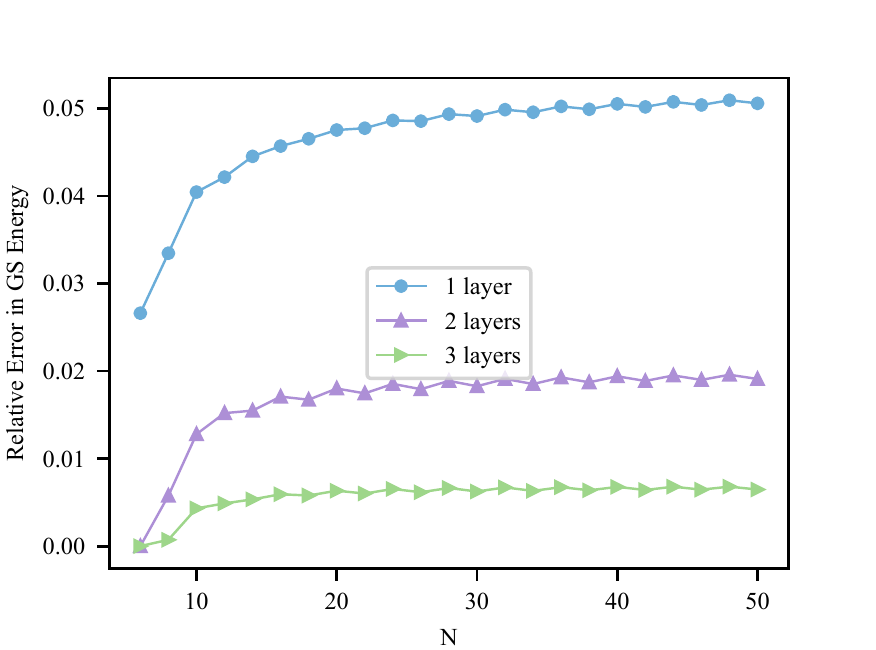}
\end{tabular}
\caption{\label{fig:FidelityLadder} \footnotesize The fidelity (a) and
  relative energy error (b) of the optimal Ansatz state and the exact
  ground state of the two-leg Heisenberg model with the total number of qubits
  $N$ using one to three layers in the Ansatz. The bond dimension we used in the MPS is $\chi=64$.}
\end{figure*}

\paragraph{Results using random parameters.} To illustrate a proof-of-principle demonstration of the potential hybrid quantum-classical approach, we have also performed additional cloud experiments on 40 qubits with random parameters in the Ansatz, and  our rZNE method gives   energy values agreeing well with the numerically calculated values; see Fig.~\ref{fig:zne102QB}ef.  This demonstrates that it is feasible to use quantum devices to extract mitigated expectation values accurately, and,  based on these,   estimate the next iteration of the variational parameters by classical computers. Hence, there is no need to know the optimal variational parameters in advance, and the rZNE-mitigated variational algorithm can potentially become practical for large-scale NISQ devices.

\paragraph{Results for the XXZ model.} As constructed, our approach works equally well for the XXZ model, and  in Fig.~\ref{fig:xxzResultBroolyn}, we present the two  sets of cloud experimental results  for a wide range of $\Delta \in[-0.8,1.4]$ with $N=40$ and $N=80$ spins, respectively, carried out on two separate backends,  {\tt ibmq\_brooklyn}  and {\tt ibm\_washington}. The mitigated values agree well with the anticipated Ansatz values. (The results for additional experiments  with $N=8$ chain are shown in Fig.~\ref{fig:xxzResult}). 

\section{Beyond one dimension.}
We expect that our protocol  can be generalized to two-dimensional structures.  
As a concrete example beyond the strict 1D, we consider a two-leg ladder of the XXZ model (see 
Fig.~\ref{fig:LadderAnsatz}).
We first prepare all vertical pairs of  qubits in singlets,  and apply the XXZ Anastz gates 
to all  horizontal odd bonds,  even bonds, and then all vertical bonds. This
constitutes a one-layer Ansatz and can be repeated for multiple layers.  
We have performed numerically simulations and  found that for $N=6$ (total number of
spins in the ladder), the two-layer Ansatz can achieve the exact ground state
for $\Delta> -1$ up to machine precision.  It is likely that for larger $N$, exact ground states can be achieved by using more layers.
 Our Ansatz achieves very high GS fidelity, exceeding 0.95 even for $N=50$ with just three layers, as well as high acurracy in the GS energy; see Fig.~\ref{fig:FidelityLadder}. 
There is an interesting phase diagram from this two-leg model~\cite{roy2017detecting,sompet2021realising}, including a Haldane phase, which could potentially be implemented on a digital quantum processor.

\section{Summary.}

In this work, we have demonstrated that variational quantum algorithms with short-depth circuits could be applied to large systems of qubits, with up to 102 qubits performed on real devices.
Despite the presence of substantial noise and errors in current devices, we have been able to improve and implement efficient error mitigation schemes to deduce accurate ground-state energy from experiments on large systems, including the use of reference states for zero-noise extrapolation, i.e., the rZNE technique. 
For the specific XXZ model, we have constructed a physics-motivated ansatz and demonstrated numerically its feasibility by analyzing the ground-state fidelity and the relative energy in the ground-state energy.
Our work thus opens up the potential practical use of error mitigated VQE on large quantum computing backends for improved accuracy. One first applies our rZNE (combining readout mitigation and possibly further mitigation) to obtain the extracted observable value(s) and/or its gradients from quantum devices, then uses classical computers to search for  variational parameters to be used in the subsequent iteration of experiments with mitigation.  The procedure is iterated until the mitigated observable value(s) converge to within certain accuracy. Such an error-mitigated, rZNE VQE approach, though not yet practical for large systems in the current cloud-based setting, due to limited allocated time and long job queues, seems plausible in dedicated experiments.  Our cloud experiments using randomly chosen parameters  already demonstrated agreement with the expected ansatz energy.
To enter a regime where quantum advantage may be realized, we will likely need to go beyond
one dimension, e.g., two dimensions, where classical simulations  of quantum many-body systems become
intractable as the system size increases. Toward this goal, we have also analyzed a two-leg ladder and showed the
applicability of our ansatz.

\paragraph*{Acknowledgements.}
We thank Paul Goldbart and Dominik Schneble for comments on the manuscript and We
 Joris Kattem\"olle for communicating his results on the exact Heisenberg ansatz. This work was supported by the National Science Foundation  under Grant No. PHY 1915165 (T.-C.W.), in particular, the part concerning properties of the model and the toolkit for extracting them, and by  the U. S. Department of Energy, Office of Science, National Quantum Information Science Research Centers, Co-design Center for Quantum Advantage (C2QA) under contract number DE-SC0012704 (H.Y. and T.-C.W.), in particular,  the design and analysis of the variational algorithm. This research also used resources of the Oak Ridge Leadership Computing Facility, which is a DOE Office of Science User Facility supported under Contract DE-AC05-00OR22725, and the Brookhaven National Laboratory operated IBM-Q Hub. The results presented in this work do not reflect the view of IBM and its employees.



\renewcommand\thefigure{S.\arabic{figure}}
\setcounter{figure}{0}
\renewcommand\thetable{S.\arabic{table}}
\setcounter{table}{0}



\appendix

\begin{figure*}[ht!]
\centering
\begin{tabular}{l}
(a) \\
\includegraphics[width=0.4\textwidth]{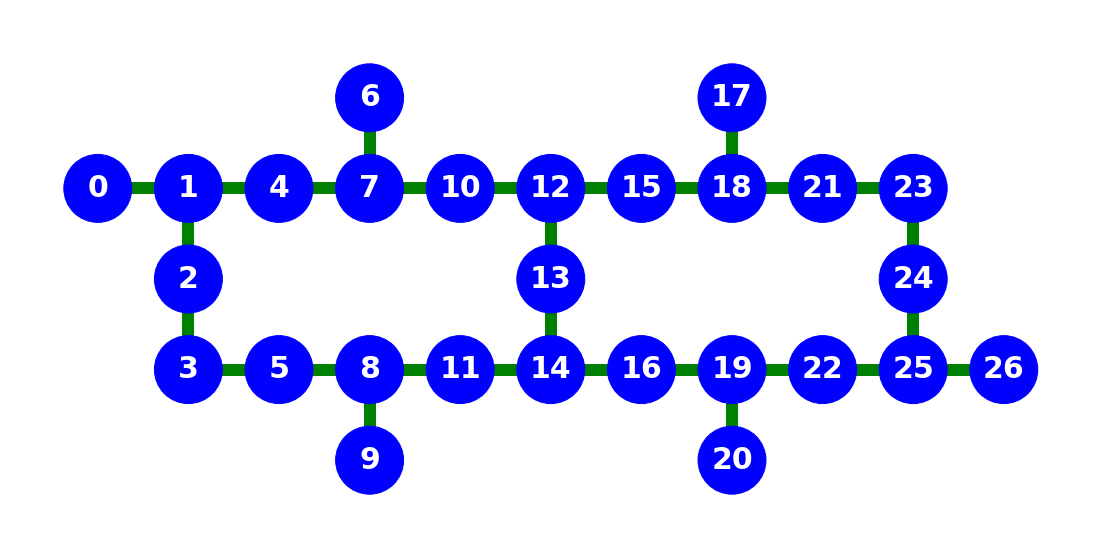}\\
(b) \\
\includegraphics[width=0.4\textwidth]{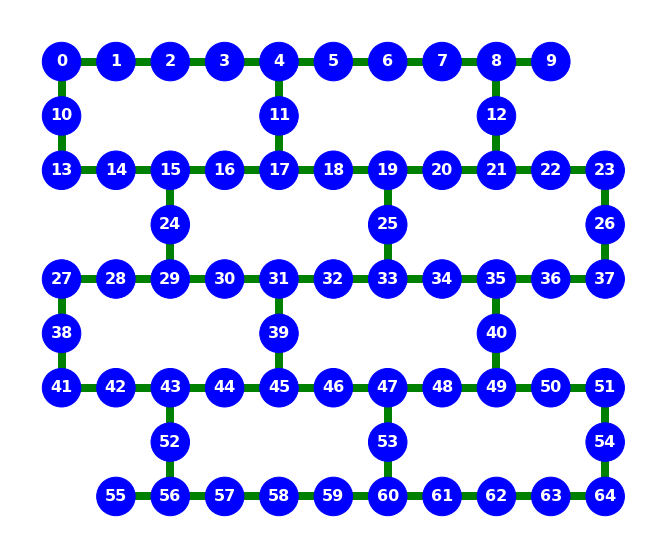}\\
\end{tabular}
\begin{tabular}{l}
(c)\\
\includegraphics[width=0.54\textwidth]{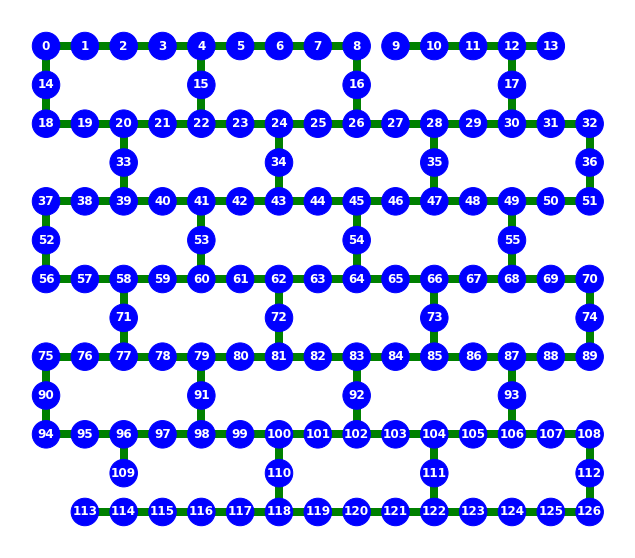}\\
\end{tabular}
\caption{\label{fig:Backends}  Illustration of the layout of some backends used in this work: (a) the layout of 27-qubit machines, such as {\tt ibm\_auckland}, {\tt ibm\_cairo}, {\tt ibm\_hanoi}, {\tt ibmq\_kolkata}, {\tt ibmq\_montreal}, {\tt ibmq\_mumbai}, and {\tt ibm\_toronto}; (b) the 65-qubit layout of  {\tt ibmq\_brooklyn}; and (c) the 127-qubit layout of {\tt ibm\_washington}. An edge between two qubits indicates that a direct CNOT gate can be executed between them. See Table~\ref{tab:backends} for certain properties of these backends.
}
\end{figure*}

\section{Properties of quantum backends and the choice of qubits} The properties of the nine quantum backends of IBM are listed in Table~\ref{tab:backends} and there are three different  layouts,  as illustrated in Fig.~\ref{fig:Backends}. Seven of the backends have 27 qubits,  the backend {\tt ibmq\_brooklyn} has 65 qubits, and {\tt ibm\_washington} has 127 qubits, with the last   also shown in Fig.~\ref{fig:FidelityLayers}b.  Before cloud experiments were performed, we examined the detailed error rates reported on the service website  and chose a path with a desired total number of sites along those connected qubits so as to avoid CNOT links with high error rates. For large system sizes, it is inevitable that we encounter a few CNOT links that may have somewhat higher error rates than others.  We note that the detailed noise and error rates may drift over time as the devices are regularly calibrated and this impact large paths more than small ones. For example, in order to perform the 80-qubit XXZ model cloud experiments in Fig.~\ref{fig:xxzResultBroolyn}b, we had to use a different path from the one used previously  for the Heisenberg model (reported in Table~\ref{tab:127}) to avoid certain CNOT links with large error rates.

\section{Additional cloud experimental results}

\begin{figure*}
\begin{tabular}{ll}
(a) & (b) \\
\includegraphics[width=0.47\textwidth]{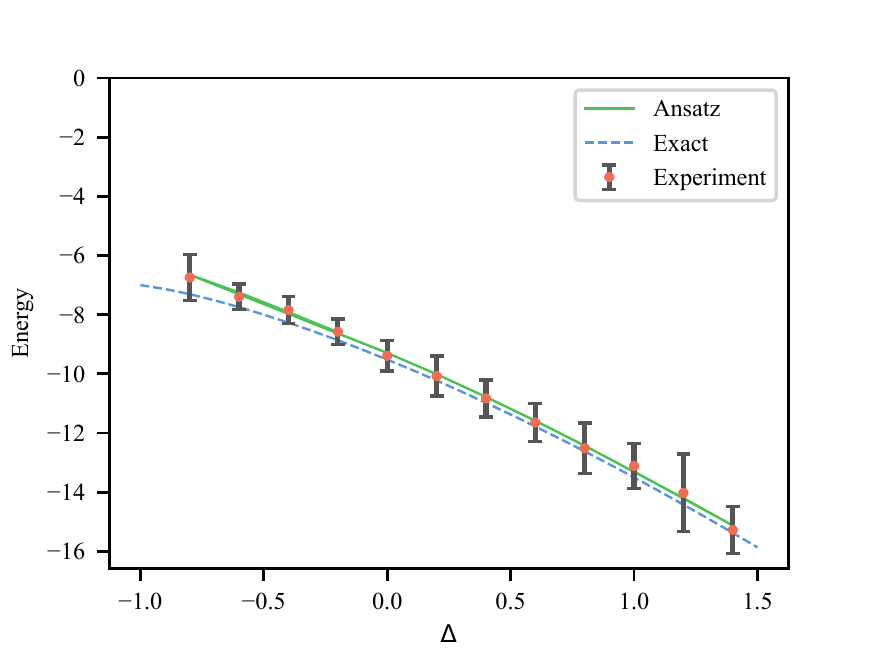}
&
\includegraphics[width=0.47\textwidth]{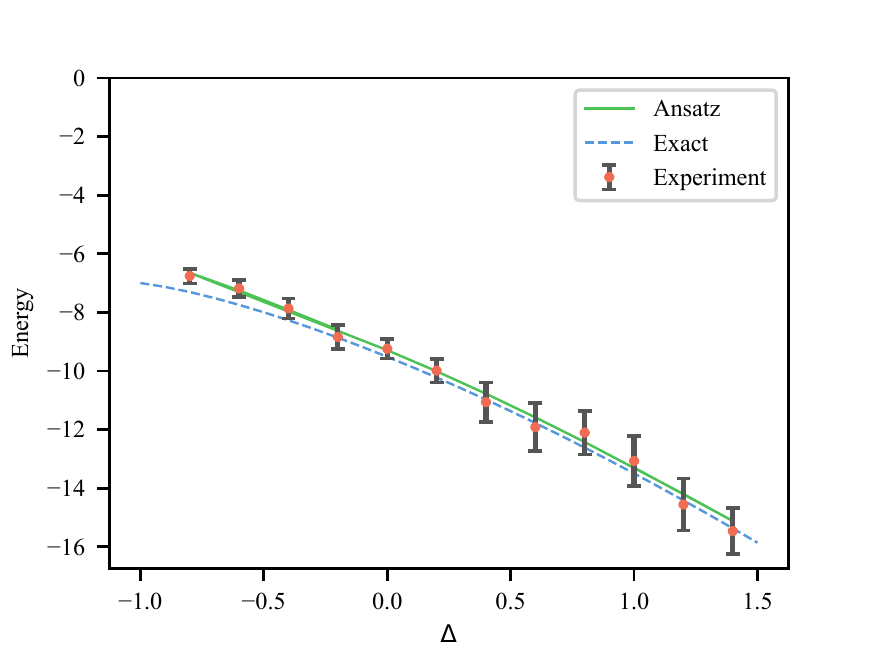}\\
\end{tabular}
\caption{\label{fig:xxzResult} The energy results for an 8-qubit XXZ chain on {\tt ibmq\_montreal}.  (a) The energy is  obtained using the Bell-measurement approach on physical qubits [15,12,13,14,16,19,22,25];  (b) The energy is  obtained using the XYZ measurement  approach on physical  qubits [11,14,16,19,22,25,24,23]. We have also performed energy measurement using quantum state tomography for $\Delta=0$ and 1.   The three methods  for measuring energy  agree very well in our cloud experiments.}
\end{figure*}

\begin{table*}[ht!]
\footnotesize
\begin{tabular}{|l|l|l|l|l|l|l|l|l|}
\hline
\begin{tabular}{l}Backend \\
(Q. volume)
\end{tabular}& $N_{\rm tot}$& Processor type  & \begin{tabular}{l}Average\\
CNOT error
\end{tabular}& 
\begin{tabular}{l}Average \\
readout error\end{tabular}  & \begin{tabular}{l}Average \\
$T_1$ time\end{tabular} & \begin{tabular}{l}Average \\
$T_2$ time\end{tabular}\\\hline
{\tt ibm\_auckland} (64) & 27 &  Falcon r5.11
& 1.042$\times 10^{-2}$
& 1.439$\times 10^{-2}$
& 178.38 $\mu$s
& 152.09 $\mu$s\\
{\tt ibmq\_brooklyn} (32) & 65  & Hummingbird r2
& 2.842$\times 10^{-2}$ &
2.928$\times 10^{-2}$ &
 74.35 $\mu$s
&  77.66 $\mu$s\\
{\tt ibm\_cairo} (64) & 27 & Falcon r5.11
& 7.969$\times 10^{-2}$
& 1.352$\times 10^{-2}$
& 101.71 $\mu$s
& 132.51 $\mu$s\\
{\tt ibm\_hanoi} (64) & 27 & Falcon r5.11
& 4.444$\times 10^{-2}$
& 1.357$\times 10^{-2}$
&151.26 $\mu$s
& 116.79 $\mu$s\\
{\tt ibmq\_kolkata} (128) & 27  & Falcon r5.11
& 4.801$\times 10^{-2}$
& 1.556$\times 10^{-2}$
&118.67 $\mu$s
& 96.82 $\mu$s\\
{\tt ibmq\_montreal} (128) & 27  &  Falcon r4
& 1.943$\times 10^{-2}$
&3.426$\times 10^{-2}$
& 119.39 $\mu$s
& 102.78 $\mu$s\\
{\tt ibmq\_mumbai} (128) & 27 &  Falcon r5.1
& 7.984$\times 10^{-2}$
& 2.665$\times 10^{-2}$
& 135.78 $\mu$s
& 117.56 $\mu$s\\
{\tt ibmq\_toronto} (32) & 27 &  Falcon r4
&  8.680$\times 10^{-2}$
& 6.050$\times 10^{-2}$
&115.84 $\mu$s
& 104.92 $\mu$s
\\
{\tt ibm\_washington} (64) & 127 &  Eagle r1 & 4.734$\times 10^{-2}$ & 2.789$\times 10^{-2}$ &  94.38 $\mu$s &
90.82 $\mu$s\\\hline
\end{tabular}
\caption{\footnotesize Properties of various IBM Q backends used in this work. Q. volume is the quantum volume and $N_{\rm tot}$ is the total number of qubits in the backend. The basis gate set of these backends include CX, ID, RZ, SX, and X, where CX denotes the CNOT gate, ID is the identity gate, RZ is the z-rotation gate, and SX is the square root of the Pauli X  gate.\label{tab:backends}}
\end{table*}

\begin{figure*}
\begin{tabular}{ll}
(a) & (b) \\
\includegraphics[width=0.48\textwidth]{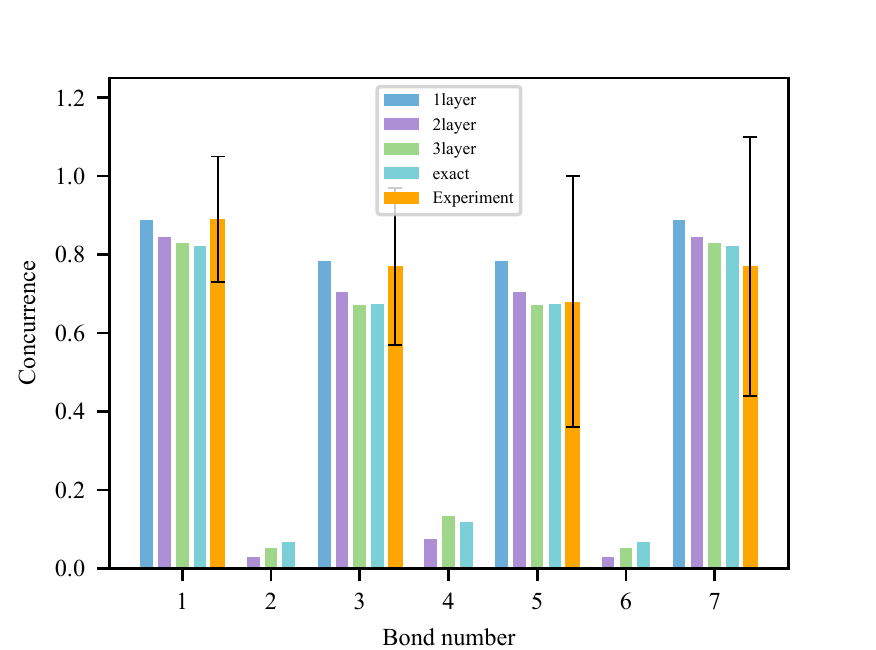} 
&
\includegraphics[width=0.48\textwidth]{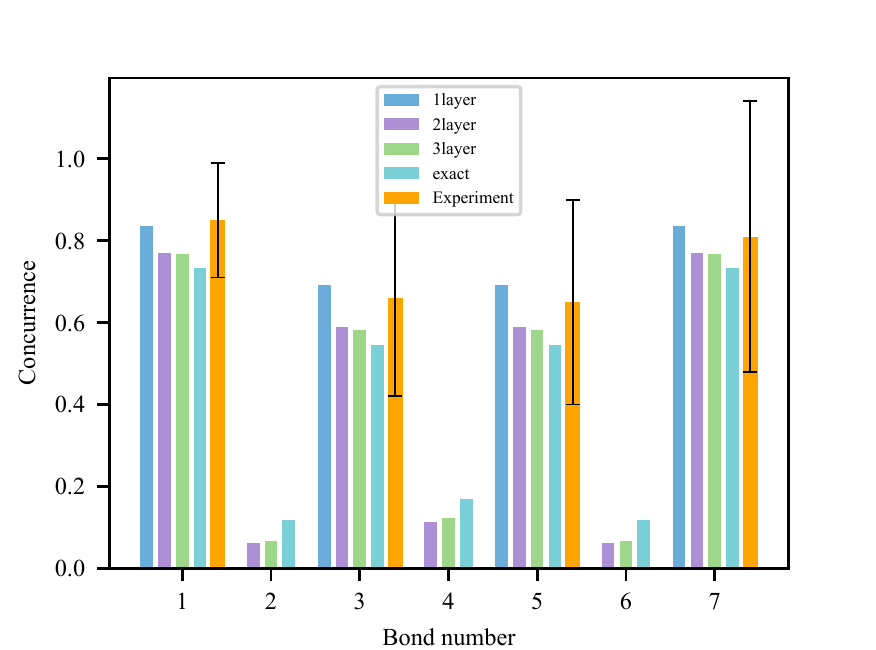}
\end{tabular}
\caption{\label{fig:ConcurrenceN8}   Ground-state entanglement property---concurrence---for a chain of 8-spin XXZ model with the open-boundary condition: 
(a) $\Delta=1$ and (b) $\Delta=0$. The concurrence is calculated for   two neighboring qubits $(j,j+1)$ on $j$-th bond ($j\in[1,7]$) using quantum states obtained from optimizing one-layer, two-layer, and three-layer Ansatzes, as well as from exact diagonalization of the XXZ Hamiltonian and from the cloud experiment  done on {\tt ibmq\_montreal} with the one-layer Ansatz. Note that with just one layer, the concurrence on the even bonds is zero. 
From these we observe that the entanglement is decreasing from the 1-layer optimal Ansatz to 2- and to  the 3-layer one, towards the exact solution. The reason is that the initial state of the Ansatz is a product of singlet Bell states on odd bonds, which possesses a very high global entanglement. The gates on even bonds act to decrease the entanglement of Bell states (on odd bonds) to increase the entanglement on  even bonds. 
}
\end{figure*}

\smallskip \noindent{\bf  Results on Heisenberg chains}. The nine IBM Q backends we use have three different layouts, as illustrated in Fig.~\ref{fig:Backends}. The complete list of the results from the cloud experiments for the Heisenberg model on various backends  and with various number of qubits is shown in Tables~\ref{tab:27and65} and~\ref{tab:127}.  These were carried out using the Bell-measurement approach.  

\smallskip \noindent{\bf Results on XXZ chains}. We have also performed cloud experiments for 8-qubit XXZ model on {\tt ibmq\_montreal}, with $\Delta$ ranging from -0.8 to 1.4, and use two different measurement methods to calculate the energy, as shown in the Fig.~\ref{fig:xxzResult}. The two methods of the XYZ measurement and of the Bell measurement agree with each other. In addition, we have also used quantum state tomography to measure the total energy at two different values of $\Delta$ (0 and 1); the energy values obtained from tomography also agree with the other two approaches. In particular, the energy results from the state tomography give $-13.46\pm0.31$ at $\Delta=1$ and $-9.3\pm0.8$ at $\Delta=0$ .

\smallskip\noindent \textbf{Concurrence results}. With the tomography approach, we have obtained additionally the concurrence for all the bonds, and the cloud experimental results are compared to those of the Ansatzes and the exact solution in Fig.~\ref{fig:ConcurrenceN8}.  Due to the open boundary condition, the concurrence alternates from large to small between odd and even bonds. The entanglement on all even bonds is identically zero for the one-layer Ansatz. This is due to the initial state being product of singlet pairs on odd bonds and the one-layer entangling operation on even bonds is not strong enough to make the pairs on even bonds entangled. 
For odd bonds, the concurrence values inferred from the clound experiments are \{0.890372, 0.767076, 0.683096, 0.768255\} at $\Delta=1$ and \{0.850059, 0.663988, 0.648982, 0.812279\} at $\Delta=0$.  
As the quantum phase transition at the Heisenberg point $\Delta=1$ is infinite-order, the concurrence does not exhibit singularity across the transition, so we did not perform cloud experiments for the concurrence over a wide of $\Delta$, but only for $\Delta=0 \& 1$ as an illustration. These concurrence values were obtained by use our rZNE approach with the naive extrapolation using Bell pairs as the reference. In doing ZNE, we had to repeat $(\mathbf{U}\mathbf{U}^{-1})$ several times, but the resulting reduced density matrices  become unentanlged for $n\ge 2$ and this makes a fitting not possible. The error bar is thus not directly accessible, but can be estimated from the energy curves.

\smallskip\noindent{\bf Information for Fig.~\ref{fig:xxzResultBroolyn}  in the main text}.  (a) A 40-qubit XXZ chain on {\tt ibmq\_brooklyn} with physical qubits  being [38,41,42,43,52,56,57,58,59,60,53,47,46,45,39,31, 30,29,24,15,16,17,11,4,5,6,7,8,12,21,20,19,25,33,34, 35,40,49,50,51]. (b) An 80-qubit XXZ chain on {\tt ibm\_washington}, with physical qubits being  [97,96,95,94,90,75,76,77,71,58,57,56,52,37,38,39, 33,20,21,22,23,24,25,26,27,28,29,30,31,32,36,51,50, 49,48,47,46,45,44,43,42,41,53,60,61,62,63,64,65,66, 67,68,69,70,74,89,88,87,93,106,107,108,112,126, 125,124,123,122,111,104,103,102,101,100,110,118, 117,116,115,114].


\section{Quantum observable depth}

\begin{figure}
\centering
\includegraphics[width=0.5\textwidth]{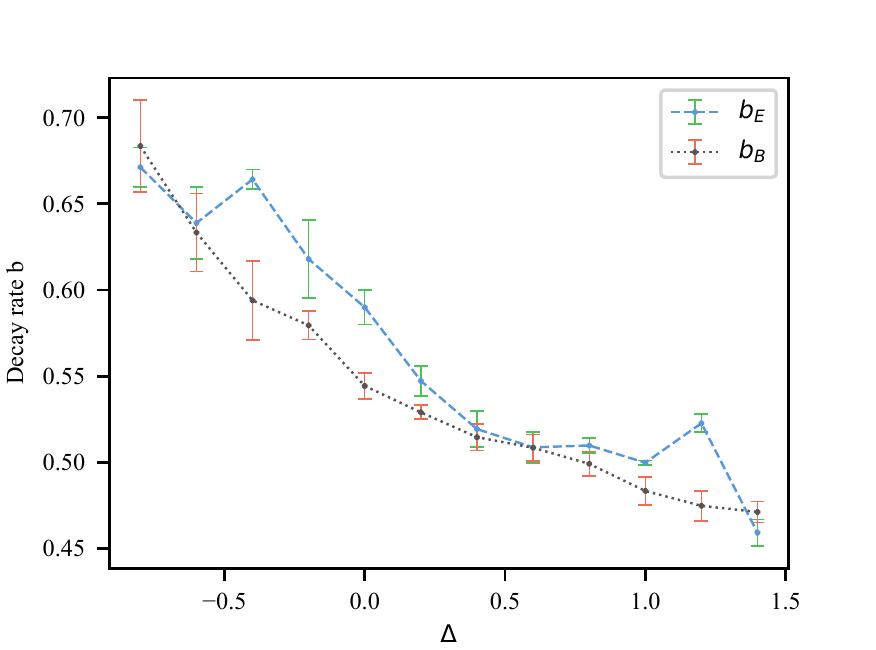}
\caption{\label{fig:BvsDelta} \footnotesize The decay coefficient $b$ from extracting  the Ansatz energy ($b_E$) and from the Bell pairs energy ($b_B$) vs. the anisotropy parameter $\Delta$ in the 80-qubit XXZ model, performed in {\tt ibm\_washington}. The QOD is related to the decay rate $b$ as ${\rm QOD}=7/b$. }
\end{figure}
In fitting the energy data, we use an exponential function $f_E(m)= a \exp(- b m) +c$, where $m=2n+1$ is the total number of $\mathbf{U}$ or $\mathbf{U}^{-1}$ in the circuit to construct the state. We note that each $\mathbf{U}$ contains 7 layers of CNOT gates. When such an exponential-decay fitting works, the quantity $7/b$, roughly speaking, represents the decay depth in the quantum circuit for the total energy, which we will refer to as the quantum observable depth (QOD), with the observable being the total energy here. It basically provides a practical way to measure how the experimental observable value degrades with the number of CNOT layers (as CNOT gates have the largest error rates in basis gate set).  From the Ansatz energy data of the 102-qubit cloud experiment on {\tt ibm\_washington}, we obtain its $b$ parameter to be $b_{E}=0.567\pm 0.03$ and hence about $12.3\pm 0.7$ value of the QOD. For the Bell pairs data, we extract that its $b$ parameter to be $b_B=0.53\pm0.05$ and hence a value of $13.1\pm 1.2$ for the QOD.  These two values seem to agree and we average them to yield a QOD of $12.7\pm0.7$. (The other set of 102-qubit cloud experiments gives a QOD of $12.59\pm0.34$.) 
The QOD depends on the qubits used in the cloud experiment and possibly on the number of qubits as well. The 50-qubit cloud experiments on {\tt ibmq\_brooklyn} give a QOD of $18.7(1.5)$. Among all the cloud experiments carried out on the backend {\tt ibm\_washington}, we find the cloud experiment using the 10 qubits [30,31,32,36,51,50,49,48,47,35] gives the best QOD value of $44\pm7$.  For the QOD from other cloud experiments and other backends, see Tables~\ref{tab:127} and~\ref{tab:27and65}.
The QOD serves as a quality measure of those qubits in the quantum processor involved in the benchmark, analogous to but  different from the metrics, such as  the randomized benchmarking and  the quantum volume. We note the QOD will depend on the choice of the observable and the model used, in particular, its value varies across different values of $\Delta$ in the XXZ model; see e.g. the decay coefficient $b$ extracted for the 80-qubit XXZ model in Fig.~\ref{fig:BvsDelta}. 
Moreover, the form of the fitting function may be different; e.g. for some prior cloud experiments with small number of qubits, both linear and quadratic fits were used in the CNOT-gate mitigation~\cite{Dumitrescu,Klco,Urbanek}. In these cases, we may need to use other quantities (such as the slope) to define the notion similar to the QOD.

\begin{table*}
\scriptsize
\begin{center}
\begin{tabular}{|c|c|c|c|c|c|c|c|}
   \hline
  $N$ & $E_{\rm exp}$ & $\epsilon_{\rm ans}$ & $\epsilon_{\rm gs}$   & QOD & qubits used & shots & rep.\\
   \hline \hline
     {10}&-16.9(3.4)  & 1.08\% & 0.78\%  & 44(7) &[30,31,32,36,51,50,49,48,47,35] &40K &50\\
    10 &
  -16.5(1.4) & 1.31\% &3.12\% & 31(4) &[32,36,51,50,49,48,47,35,28,29] &40K & 25
   \\
    20 &-33.8(1.4) &0.055\%&2.67\% & 20.3(1.0) & \begin{tabular}{c}
    [50,51,36,32,31,30,29,28,35,47,\\
48,49,55,68,69,70,74,89,88,87]\end{tabular}& 40K & 25\\
    30 & -51.1(2.3)& 0.36\%& 2.57\% &  19.6(1.1) &\begin{tabular}{c}
   [115,116,117,118,110,100,101,102,\\
   92,83,84,85,73,66,67,68,55,49,50,\\
   51,36,32,31,30,29,28,27,26,16,8] 
   \end{tabular}& 40K & 50\\
    40 & -69.2(1.6)& 1.74\% &1.38 & 12.1(0.8) &\begin{tabular}{c}[2,1,0,14,18,19,20,33,39,38,37,52,56,57,\\
    58,59,60,61,62,72,81,82,83,92,102,103,104,\\
    105,106,93,87,86,85,73,66,67,68,55,49,48]\end{tabular} & 40K & 50\\
    50 &-86.9(1.8) & 2.09\%& 1.12\% & 13.19(0.28) &
 list of 40 qubits + [47,35,28,27,26,25,24,23,22,15]  & 40K & 25\\
    60 &-99(4) & 3.15\% & 6.26\% & 11.28(0.34) &\begin{tabular}{c}
   [3,2,1,0,14,18,19,20,33,39,38,37,52,56,57,58,59,\\
   60,53,41,42,43,34,24,25,26,27,28,29,30,31,32,36,\\
   51,50,49,48,47,46,45,54,64,63,62,72,81,82,83,92,\\
   102,103,104,105,106,107,108,112,126,125,124]\end{tabular} & 40K & 50\\
    70 &-125(7) &4.76\% & 1.35\% & 13.1(0.8) &\begin{tabular}{c}
   [3,2,1,0,14,18,19,20,21,22,23,24,25,26,27,28,29,30,31,32,\\
   36,51,50,49,48,47,46,45,54,64,65,66,67,68,69,70,74,89,\\
   88,87,93,106,105,104,111,122,121,120,119,118,110,100,\\
   101,102,92,83,82,81,80,79,91,98,97,96,95,94,90,75,76,77] \end{tabular}&40K & 50\\
    80 & -138.5(2.5)& 1.52\% & 1.82\% & 12.79(0.23) &\begin{tabular}{c}[3,2,1,0,14,18,19,20,33,39,38,37,52,56,57,58,59,60,\\
    53,41,42,43,34,24,25,26,27,28,29,30,31,32,36,51,50,\\
    49,55,68,69,70,74,89,88,87,93,106,105,104,111,122,\\
    121,120,119,118,110,100,101,102,92,83,84,85,73,66,\\
    65,64,63,62,72,81,80,79,91,98,97,96,95,94,90,75]\end{tabular}& 40K & 50\\
    90 & -153(5)& 0.34\% & 3.64\%& 12.5(0.4) &\begin{tabular}{c}
   [3,2,1,0,14,18,19,20,21,22,23,24,25,26,27,28,29,\\
   30,31,32,36,51,50,49,48,47,46,45,44,43,42,41,40,\\
   39,38,37,52,56,57,58,59,60,61,62,63,64,65,66,67,\\
   68,69,70,74,89,88,87,86,85,84,83,82,81,80,79,78,\\
   77,76,75,90,94,95,96,97,98,99,
100,101,102,103,\\
104,105,106,107,108,112,126,125,124,123,122]\end{tabular} & 40K & 25\\
    98 & -168.1(2.6) & 0.54\% &2.81\% &12.32(0.27) &\begin{tabular}{c} [3,2,1,0,14,18,19,20,33,39,40,41,42,43,34,\\
    24,23,22,15,4,5,6,7,8,16,26,27,28,29,30,31,\\
    32,36,51,50,49,48,47,46,45,54,64,63,62,61,60,59,58,\\
    71,77,76,75,90,94,95,96,97,98,91,79,80,81,82,\\
    83,84,85,73,66,67,68,69,70,74,89,88,87,93,106,\\
    107,108,112,126,125,124,123,122,111,\\
    104,103,102,101,100,110,118,117,116,115,114]\end{tabular} & 40K & 75\\
    100 & -173(9) &1.39\% & 1.99\% & 11.9(0.14) &\begin{tabular}{c}[3,2,1,0,14,18,19,20,21,22,15,4,5,6,7,8,16,26,\\
    27,28,29,30,31,32,36,51,50,49,48,47,46,45,44,\\
    43,42,41,40,39,38,37,52,56,57,58,59,60,61,62,\\
    63,64,65,66,67,68,69,70,74,89,88,87,86,85,84,\\
    83,82,81,80,
79,78,77,76,75,90,94,95,
96,97,98,\\
99,100,101,102,103,104,105,106,107,108,112,\\
126,125,124,123,122,121,120,119,118,117,116]\end{tabular} & 40K & 50\\
    102 &-177.5(2.7) & 1.99\% & 1.42\% & 12.23(0.17) & list of 100 qubits +[115,114] & 40K & 75\\
   \hline  
\end{tabular}
\end{center}
\caption{\footnotesize Various Heisenberg spin-chain cloud experiments performed on the 127-qubit  {\tt ibm\_washington} backend/device of IBM Q. }
\label{tab:127}
\end{table*}
\begin{table*}[h!]
\scriptsize
\begin{center}
\begin{tabular}{|c|c|c|c|c|c|c|c|}
\cline{1-2}
\multicolumn{2}{|l|}{ Backend}\\
    \hline
 
  $N$ & $E_{\rm exp}$ & $\epsilon_{\rm ans}$ & $\epsilon_{\rm gs}$   & QOD & qubits used & shots & rep.\\
   \hline\hline
    \multicolumn{2}{|l|}{ ibm\_auckland}\\
    \hline
 20 & -34.3(1.3)  & 1.42\% & 1.24\%& 13.4(0.5) &{ [13,12,10,7,4,1,2,3,5,8,11,14,16,19,22,25,24,23,21,18]} & 100K & 50\\\hline
      \multicolumn{2}{|l|}{ ibm\_cairo}\\
    \hline
     16 & -27.5(2.5) & 1.93\% & 0.53\% & 13.5(1.1) &{[1,2,3,5,8,11,14,16,19,22,25,24,23,18,15,12]}& 100K & 50\\
      18 & -28.9(1.3) & 4.93\% & 7.34\% & 13.4(0.4) &{[1,2,3,5,8,11,14,16,19,22,25,24,23,18,15,12]}& 100K & 50\\
     18 & -32.3(1.1) & 6.25\% & 3.57\% & 13.51(1.1) &{[6,7,4,1,2,3,5,8,11,14,16,19,22,25,24,23,21,18]}& 100K & 50\\
     20 &-32.8(1.1) & 3.01\% & 5.56\% & 12.47(0.35) &{[6,7,4,1,2,3,5,8,11,14,16,19,22,25,24,23,21,18,15,12]}& 100K & 50\\\hline
        \multicolumn{2}{|l|}{ ibm\_hanoi }\\
    \hline   20 &-32.3(1.3) &4.49\% & 7.00\%& 17.0(0.7) &{[5,3,2,1,4,7,10,12,15,18,21,23,24,25,22,19,16,14,11,8]}& 100K & 100\\\hline 
      \multicolumn{2}{|l|}{  ibmq\_kolkata }\\
    \hline
      20 & -33.5(1.2) & 0.94\% & 3.54\% & 18.6(0.8) &{[26,25,22,19,16,14,11,8,5,3,2,1,4,7,10,12,15,18,21,23]}& 40K & 25\\\hline 
       \multicolumn{2}{|l|}{ ibmq\_montreal }\\
    \hline 
     6 &-9.9(1.6) & 0.75\% & 0.19\% & 13.1(0.6) &{[20,19,22,25,24,23]}& 32K & 100\\
     20 &-31.7(1.7) & 6.26\% & 8.72\%& 13.0(0.5)  &{[12,10,7,4,1,2,3,5,8,11,14,16,19,22,25,24,23,21,18,17]}& 32K & 50\\\hline
     \multicolumn{2}{|l|}{ ibmq\_mumbai }\\
    \hline 
     4 &-6.5(1.6) & 0.56\% & 0.56\%& 37(9) &{[12,15,18,17]}&8192 & 146\\
     6 &-9.9(3.5) & 0.19\%&  0.75\% & 64(11) &{[7,10,12,15,18,17]}&8192 & 146\\
     8 &-13.2(2.2) & 0.75\%&2.22\% & 36(5) &{[1,4,7,10,12,15,18,21]}&8192 & 146\\
     10 &-16.6(3.3) &0.71\% &2.54\%& 38(7) &{[0,1,4,7,10,12,15,18,21,23]} & 8192 &100\\
     12 & -20.3(2.1)&0.80\% & 1.30\% & 14.0(1.4) &{[8,5,3,2,1,4,7,10,12,15,18,21]}& 8192 & 100\\
     14 & -23.6(1.8)& 0.17\% & 2.10\% & 14.1(1.0) &{[8,5,3,2,1,4,7,10,12,15,18,21,23,24]}& 8192 & 100\\
     16 &-24.0(2.1) &10.1\% & 7.42\% & 12.1(0.8) &{[8,5,3,2,1,4,7,10,12,15,18,21,23,24,25,22]}& 8192 & 100\\
     18 & -29.7(1.3)& 2.30\% & 4.77\% & 16.7(0.7) &{[8,5,3,2,1,4,7,10,12,15,18,21,23,24,25,22,19,16]}& 8192 &100\\
    20 & -31.6(2.3)& 6.56\% & 9.01\% & 14.2(1.0) &{[8,5,3,2,1,4,7,10,12,15,18,21,23,24,25,22,19,16,14,11]}& 8192 & 100\\\hline
     \multicolumn{2}{|l|}{ibm\_toronto}\\
    \hline 
    18 & -32.1(1.7)&5.60\% & 2.92\% & 13.1(0.6) &{[0,1,4,7,10,12,15,18,21,23,24,25,22,19,16,14,11,8]}& 16K & 50\\ &-32.3(1.4) & 4.49\% & 7.00\% & 13.0(0.5) &{[0,1,4,7,10,12,15,18,21,23,24,25,22,19,16,14,11,8,5,3]}& 32K & 50\\
   \hline 
    \multicolumn{2}{|l|}{ibmq\_brooklyn}\\
    \hline  
    10 & -16.8(1.6) & 0.48\% & 1.36\% & 24.0(2.1) & {[53,47,48,49,40,35,34,33,25,19]}& 100K & 50\\
  20 & -33.4(0.9) & 1.24\% & 3.83\% & 22.7(0.7) & {[9,8,7,6,5,4,3,2,1,0,10,13,14,15,16,17,18,19,25,33]}& 100K & 50\\
   20 & -34.0(3.1) & 0.54\%& 2.10\%   &  27.4(3.4) & {[0,1,2,3,4,5,6,7,8,12,21,20,19,18,17,16,15,24,29,28]}& 20K & 80\\
   30 & -49.3(3.3)& 3.18\% & 6.00\% & 18.4(1.1) & \begin{tabular}{c}[43,52,56,57,58,59,60,53,47,48,49,40,35,34,\\
 33,32,31,30,2,24,15,16,17,11,4,5,6,7,8,9]
\end{tabular} & 100K & 50\\
   40 &  -68(4)&0.028\% &3.09\%& 20.6(1.3) & \begin{tabular}{c}[63,62,61,60,53,47,46,45,39,31,32,33,25,19,\\
 18,17,16,15,14,13,10,0,1,2,3,4,5,6,7,8,\\
 12,21,22,23,26,37,36,35,40,49]
  \end{tabular} & 100K & 50\\
   50 & -83(5)& 2.49\% & 5.56\% &  18.7(1.5) &\begin{tabular}{c}
[43,52,56,57,58,59,60,61,62,63,64,54,51,50,49,\\
40,35,36,37,26,23,22,21,12,8,7,6,5,4,3,2,1,0,10,\\13,14,15,16,17,18,19,25,33,32,31,39,45,46,47,48]  \end{tabular}& 100K &50\\
  \hline
\end{tabular}
\end{center}
\caption{\footnotesize Heisenberg spin-chain cloud experiments performed on all available 27-qubit backends/devices and the 65-qubit {\tt ibmq\_brooklyn}  of IBM Q. }
\label{tab:27and65}
\end{table*}

\begin{thebibliography}{99}
\bibitem{benioff1980}
P. Benioff, \href{https://link.springer.com/article/10.1007/BF01011339}{ The computer as a physical system: A microscopic quantum mechanical Hamiltonian model of computers as represented by Turing machines. Journal of Statistical Physics, {\bf 22}, 563–591 (1980).}

\bibitem{mannin1980}
Y. I. Manin,   Vychislimoe i nevychislimoe [Computable and Noncomputable] (in Russian). Sovetskoe Radio (1980).
\bibitem{feynman1982}
R. P. Feynman,   \href{https://link.springer.com/article/10.1007/BF02650179}{Simulating physics with computers. International Journal of Theoretical Physics, {\bf 21}, 467–488 (1982).}
\bibitem{lloyd1996}
S.  Lloyd, \href{https://www.science.org/doi/abs/10.1126/science.273.5278.1073}{ Universal Quantum Simulators.
Science
{\bf  273},
1073-1078 (1996).}

\bibitem{shor1994}
P. W. Shor,  \href{https://ieeexplore.ieee.org/document/365700}{ Algorithms for quantum computation: Discrete logarithms and factoring. In Proceedings 35th Annual Symposium on Foundations of Computer Science (pp. 124--134). IEEE (1994).}

\bibitem{NISQ}
 J. Preskill,  \href{https://quantum-journal.org/papers/q-2018-08-06-79/}{ Quantum Computing in the NISQ era and beyond. Quantum {\bf 2}, 79 (2018).}


\bibitem{Endo}
 S. Endo,  S. C. Benjamin,  Y.  Li, \href{https://journals.aps.org/prx/abstract/10.1103/PhysRevX.8.031027}{  Practical Quantum Error Mitigation for Near-Future Applications. Phys. Rev. X {\bf 8}, 031027 (2018).}
\bibitem{Temme}
 K. Temme, S.  Bravyi, J. M. Gambetta, \href{https://journals.aps.org/prl/abstract/10.1103/PhysRevLett.119.180509}{ Error Mitigation for Short-Depth Quantum Circuits.
Phys. Rev. Lett. {\bf 119}, 180509 (2017).}
\bibitem{znekandala} A. Kandala, K.  Temme, A. D.  C\'orcoles, A.   Mezzacapo, J. M.  Chow, J. M.  Gambetta,  \href{https://www.nature.com/articles/s41586-019-1040-7}{ Error mitigation extends the computational reach of a noisy quantum processor. Nature, {\bf 567}, 491–495 (2019). }

\bibitem{Dumitrescu}
E. F. Dumitrescu, A. J. McCaskey, G.  Hagen, G. R.  Jansen, T. D. Morris,  T. Papenbrock, R. C. Pooser, D. J.  Dean, P. Lougovski, \href{https://journals.aps.org/prl/abstract/10.1103/PhysRevLett.120.210501}{ Cloud Quantum Computing of an Atomic Nucleus.
Phys. Rev. Lett. {\bf 120}, 210501 (2018).}
\bibitem{Klco}
N. Klco, E. F. Dumitrescu, A. J.  McCaskey, T. D. Morris, R. C.  Pooser, M.  Sanz, E. Solano, P. Lougovski, M. J.  Savage, \href{https://journals.aps.org/pra/abstract/10.1103/PhysRevA.98.032331}{ Quantum-classical computation of Schwinger model dynamics using quantum computers. Phys. Rev. A {\bf 98}, 032331 (2018).}
\bibitem{Urbanek}
M. Urbanek, B. Nachman, V. R.  Pascuzzi, A. He, C. W. Bauer,  W. A. de Jong,  
\href{https://journals.aps.org/prl/abstract/10.1103/PhysRevLett.127.270502}{Mitigating Depolarizing Noise on Quantum Computers with Noise-Estimation Circuits.
Phys. Rev. Lett. {\bf 127}, 270502 (2021).}

\bibitem{Giurgica-Tiron}
T. Giurgica-Tiron, Y.  Hindy, R.  LaRose, A. Mari,  W. J. Zeng, 
\href{http://arXiv.org/abs/2005.10921}{Digital zero noise extrapolation for quantum error mitigation. Preprint at http://arXiv.org/abs/2005.10921 (2020).}

\bibitem{wallman2016}
J. J. Wallman, J.  Emerson,  \href{https://journals.aps.org/pra/abstract/10.1103/PhysRevA.94.052325}{Noise tailoring for scalable quantum computation via randomized compiling. Phys. Rev. A {\bf 94}, 052325 (2016).}

 \bibitem{kim2021scalable}
Y. Kim, C. J. Wood, T. J. Yoder, S. T. Merkel, J. M. Gambetta, K. Temme, and A. Kandala, \href{ https://arxiv.org/abs/2108.09197}{Scalable error mitigation for noisy quantum circuits produces competitive expectation values. Preprint at https://arxiv.org/abs/2108.09197 (2021).}

\bibitem{NISQRMP}  
K. Bharti, A.  Cervera-Lierta, T. H.  Kyaw,  T.   Haug, S.  Alperin-Lea, A.  Anand, M.  Degroote, H. Heimonen,  J. S. Kottmann, T.  Menke, W.-K. Mok, S.  Sim, L.-C.  Kwek, A.  Aspuru-Guzik,  \href{https://journals.aps.org/rmp/abstract/10.1103/RevModPhys.94.015004}{Noisy intermediate-scale quantum algorithms. Rev. Mod. Phys. {\bf94}, 015004 (2022).}



\bibitem{Preskill2012}
J. Preskill,  \href{https://arxiv.org/abs/1203.5813}{Quantum computing and the entanglement frontier, Preprint at
https://arxiv.org/abs/1203.5813 (2012).}

\bibitem{GoogleSupremacy} 
F.  Arute et al.
\href{https://www.nature.com/articles/s41586-019-1666-5}{Quantum supremacy using a programmable superconducting processor.
Nature  {\bf 574}, 505-510 (2019).}

\bibitem{Zuchongzhi}
 Y. Wu  et al.
 \href{https://journals.aps.org/prl/abstract/10.1103/PhysRevLett.127.180501}{Strong Quantum Computational Advantage Using a Superconducting Quantum Processor.
Phys. Rev. Lett. {\bf 127}, 180501 (2021).}


\bibitem{Jiuzhang}

 H.-S. Zhong et al.
 \href{https://www.science.org/doi/full/10.1126/science.abe8770}{Quantum computational advantage using photons. Science {\bf 370}, 1460-1463 (2020).}
\bibitem{madsen2022quantum}
 L. S. Madsen et al.
 \href{https://www.nature.com/articles/s41586-022-04725-x}{Quantum computational advantage with a programmable photonic processor.
Nature {\bf 606}, 75–81 (2022).}

\bibitem{GoogleHartreeFock} 
F. Arute  et al.
\href{https://www.science.org/doi/full/10.1126/science.abb9811}{Hartree-Fock on a superconducting qubit quantum computer. Science {\bf  369},
1084-1089 (2020).}
\bibitem{GoogleTopological} 

K. J. Satzinger et al.
\href{https://www.science.org/doi/full/10.1126/science.abi8378}{Realizing topologically ordered states on a quantum processor. Science {\bf 374}, 1237–1241 (2021).}
\bibitem{PanQW} M. Gong et al.
\href{https://www.science.org/doi/full/10.1126/science.abg7812}{Quantum walks on a programmable two-dimensional 62-qubit superconducting processor.
Science {\bf 372}, 948-952 (2021)}

\bibitem{Heisenberg}
W.  Heisenberg, \href{https://link.springer.com/article/10.1007/BF01328601}{ Z\"ur Theorie des Ferromagnetismus (On the theory of ferromagnetism). Zeitschrift f\"ur Physik  {\bf 49}. 619–636 (1928).}

\bibitem{Kasteleijn}
 P. W. Kasteleijn, \href{https://www.sciencedirect.com/science/article/abs/pii/S0031891452802733}{The lowest energy state of a linear antiferromagnetic chain. Physica
 {\bf 18},  104-113 (1952).}
\bibitem{KorepinBook}
V. E.  Korepin, N. M. Bogoliubov,  A. G. Izergin,  \href{https://link.springer.com/chapter/10.1007/3-540-16075-2_12}{Quantum Inverse Scattering Method and Correlation Functions, Cambridge University Press (Cambridge, UK, 1993).}


\bibitem{QAOA}
E. Farhi, J.  Goldstone, S.  Gutmann,  \href{https://arXiv.org/abs/1412.6062}{A Quantum Approximate Optimization Algorithm Applied to a Bounded Occurrence Constraint Problem. Preprint at https://arXiv.org/abs/1412.6062 (2014).}


\bibitem{kattemolle2021variational}
 J. Kattem{\"o}lle, and J. van Wezel, \href{https://arXiv.org/abs/2108.02175}{Variational quantum eigensolver for the Heisenberg antiferromagnet on the kagome lattice,
 preprint arXiv:2108.02175 (2021).}



\bibitem{VatanWilliams}
 F. Vatan, C.  Williams,  \href{https://journals.aps.org/pra/abstract/10.1103/PhysRevA.69.032315}{Optimal Quantum Circuits for General Two-Qubit Gates. Phys. Rev. A {\bf 69}, 032315 (2004).}

\bibitem{BCS}
 J. Bardeen, L. N.  Cooper, J. R.  Schrieffer, \href{https://journals.aps.org/pr/abstract/10.1103/PhysRev.108.1175}{  Theory of Superconductivity. Phys. Rev. {\bf 108} (5), 1175–1204 (1957).}
\bibitem{Laughlin}
 R. B. Laughlin,  \href{https://journals.aps.org/prl/abstract/10.1103/PhysRevLett.50.1395}{Anomalous Quantum Hall Effect: An Incompressible Quantum Fluid with Fractionally Charged Excitations. Phys. Rev. Let. {\bf 50} (18),  1395–1398 (1983).}
 

\bibitem{NielsenChuang00}
 M. Nielsen, I.  Chuang,  \href{https://www.cambridge.org/highereducation/isbn/9780511976667}{{Quantum Computation and Quantum Information\/}
(Cambridge Univ. Press, 2000).}

\bibitem{Joris}
 For larger $N$, such as 8 and 10, exact ground states can be achieved with more layers. We thank Joris Kattem\"olle for communicating his results. 
 
\bibitem{PEPS}
F. Verstraete, J. I.  Cirac,  V. Murg,  \href{https://www.tandfonline.com/doi/full/10.1080/14789940801912366}{Matrix Product States, Projected Entangled Pair States, and variational renormalization group methods for quantum spin systems.
 Adv. Phys. {\bf 57}, 143 (2008).}
\bibitem{MPS1}
 S. \"Ostlund, S. Rommer,  
\href{https://journals.aps.org/prl/abstract/10.1103/PhysRevLett.75.3537}{Thermodynamic Limit of Density Matrix Renormalization. 
Phys. Rev. Lett. {\bf 75}, 3537 (1995).}
\bibitem{MPS2}
M. Fannes, B.   Nachtergaele, R. F.  Werner, 
\href{https://link.springer.com/article/10.1007/BF02099178}{Finitely correlated states on quantum spin chains. 
Commun. Math.
Phys. {\bf 144}, 443 (1992).}


\bibitem{DMRG}
S. R. White,  \href{https://journals.aps.org/prl/abstract/10.1103/PhysRevLett.69.2863}{Density matrix formulation for quantum renormalization groups. Phys. Rev. Lett. {\bf 69}, 2863 (1992).}


 

\bibitem{keith2018joint}
 A. C. Keith, C. H.   Baldwin, S.  Glancy, E.  Knill,  \href{https://journals.aps.org/pra/abstract/10.1103/PhysRevA.98.042318}{Joint
quantum-state and measurement tomography with incomplete
measurements. Phys. Rev. A {\bf 98}, 042318 (2018).}

\bibitem{chen2019detector}  Y. Chen, M. Farahzad, S. Yoo, T.-C. Wei,   \href{https://journals.aps.org/pra/abstract/10.1103/PhysRevA.100.052315}{Detector Tomography on IBM 5-qubit Quantum Computers and Mitigation of Imperfect Measurement, Phys. Rev. A {\bf 100}, 052315 (2019).}


\bibitem{geller2021efficient}
M. R. Geller, M.  Sun,  \href{https://doi.org/10.1088/2058-9565/abd5c9}{Efficient correction of multiqubit measurement errors. Quantum Sci. Technol. {\bf 6},  025009 (2021).}
\bibitem{maciejeswski2020}
 F. B. Maciejewski, Z. Zimbor\'as, M.  Oszmaniec,  \href{https://quantum-journal.org/papers/q-2020-04-24-257/}{Mitigation of readout noise in near-term quantum devices by classical post-processing based on detector tomography. Quantum {\bf 4}, 257 (2020).}
\bibitem{CTMP} S. Bravyi, S. S. Sheldon,  A.  Kandala, D. C.  Mckay, J. M.  Gambetta,  \href{https://journals.aps.org/pra/abstract/10.1103/PhysRevA.103.042605}{Mitigating measurement errors in multi-qubit experiments. Phys. Rev. A {\bf 103}, 042605 (2021). }
\bibitem{M3Miti} P. D. Nation, H. Kang, N. Sundaresan, J. M.  Gambetta,  \href{https://arXiv.org/abs/2108.12518}{ Scalable mitigation of measurement errors on quantum computers. Preprint at https://arXiv.org/abs/2108.12518  (2021).}

\bibitem{roy2017detecting}
 S. S. Roy, H. S. Dhar, D.  Rakshit, A.  Sen(De), U. Sen, 
\href{https://www.sciencedirect.com/science/article/pii/S0304885317322412}{Detecting phase boundaries of quantum spin-1/2 XXZ ladder via bipartite and multipartite entanglement transitions.
Journal of Magnetism and Magnetic Materials {\bf 444},
 227-235 (2017).}
 
\bibitem{sompet2021realising}
P. Sompet, S.  Hirthe, D. Bourgund, T.  Chalopin, J.
 Bibo, J.  Koepsell, P.  Bojovi\'c, R.  Verresen, F. Pollmann, 
G. Salomon, C.  Gross, T. A.  Hilker, I. Bloch,   \href{https://arXiv.org/abs/2103.10421}{Realising the Symmetry-Protected Haldane Phase in Fermi-Hubbard Ladders. Preprint at https://arXiv.org/abs/2103.10421 (2021).}




\end{thebibliography}
\end{document}